\documentclass[12pt]{article}
\pdfoutput=1

\def \be {\begin{equation}}
\def \ee {\end{equation}}
\def \bea {\begin{eqnarray}}
\def \eea {\end{eqnarray}}

\newcommand{\pd}{\partial}

\newcommand{\ppsi}{p_{\psi}}
\newcommand{\pphi}{p_{\phi}}
\newcommand{\Qo}{Q_{1}}
\newcommand{\Qf}{Q_{2}}

\newcommand{\gto}{\tilde{\gamma}_1}
\newcommand{\gtt}{\tilde{\gamma}_2}
\newcommand{\mpsi}{m_{\psi}}
\newcommand{\mphi}{m_{\phi}}
\newcommand{\wb}{\tilde{\omega}}
\newcommand{\lb}{\tilde{\lambda}}

\usepackage{enumerate}
\usepackage{amsmath}
\usepackage{amssymb}
\usepackage{graphicx,subfigure}

\newtheorem{theorem}{Theorem}[section]

\title{Instability of supersymmetric microstate geometries}

\author{Felicity C. Eperon, Harvey S. Reall and Jorge E. Santos \\ {\small Department of Applied Mathematics and Theoretical Physics, University of Cambridge} \\ {\small Wilberforce Road, Cambridge CB3 0WA, UK}\\{\small fce21, hsr1000, jss55@cam.ac.uk}}

\begin{document}

\maketitle

\begin{abstract}
We investigate the classical stability of supersymmetric, asymptotically flat, microstate geometries with five non-compact dimensions. Such geometries admit an "evanescent ergosurface": a timelike hypersurface of infinite redshift. On such a surface, there are null geodesics with zero energy relative to infinity. These geodesics are stably trapped in the potential well near the ergosurface. We present a heuristic argument indicating that this feature is likely to lead to a nonlinear instability of these solutions. We argue that the precursor of such an instability can be seen in the behaviour of linear perturbations: nonlinear stability would require that all linear perturbations decay sufficiently rapidly but the stable trapping implies that some linear perturbation decay very slowly. We study this in detail for the most symmetric microstate geometries. By constructing quasinormal modes of these geometries we show that generic linear perturbations decay slower than any inverse power of time. 
\end{abstract}

\section{Introduction}

Type IIB supergravity admits supersymmetric "microstate geometry" solutions \cite{Lunin:2001jy,maldacena2000,balasubramanian2000,lunin2002,Lunin:2004uu,Giusto:2004id,Giusto:2004ip,Giusto:2004kj,Bena:2005va,Berglund:2005vb}. These are asymptotically flat, geodesically complete, stationary solutions without horizons. Near infinity, they approach the product of 5 dimensional flat spacetime with $5$ compact dimensions. Some of these solutions can be dimensionally reduced to give smooth solutions of 5d supergravity. In 5d, the stationary Killing vector field $V$ is timelike everywhere except on a certain timelike hypersurface, where is it is null. This surface has infinite redshift relative to infinity, and has been called an "evanescent ergosurface" \cite{Gibbons:2013tqa}.

A natural question is whether these spacetimes are classically stable. This has been investigated for non-supersymmetric microstate geometries, which can have a genuine ergoregion, where $V$ becomes spacelike \cite{Jejjala:2005yu}. Such geometries have been shown to be unstable \cite{Cardoso:2005gj}: linear perturbations localized in the ergoregion can have negative energy and there exist modes which grow exponentially in time. In the supersymmetric case, linear perturbations have non-negative energy, which excludes exponential growth so one might expect stability. 

A simple argument suggests that supersymmetric microstate geometries actually have a nonlinear instability. The argument is based on the existence of the evanescent ergosurface. As we shall explain, on an evanescent ergosurface, $V$ is tangent to affinely parameterized null geodesics with zero energy. These geodesics are at rest relative to infinity so they are resisting the frame-dragging effect caused by the rotation of the geometry. Hence they can be regarded as carrying angular momentum opposed to that of the background spacetime. These geodesics are "trapped" in the sense that they remain in a finite region of space, i.e., they do not disperse. Trapping occurs in other situations, e.g., at the photon sphere of a Schwarzschild black hole. However, in the Schwarzschild case, the trapping is unstable: if one perturbs the geodesic then it will escape to infinity or fall into the black hole. At an evanescent ergosurface the trapping is {\it stable} because the geodesics sit at the bottom of a gravitational potential well.

Now consider perturbing the spacetime by adding an uncharged massive particle (or a tiny black hole) near to the evanescent ergosurface. If we neglect backreaction then the particle moves on a geodesic. However, if we couple it to supergravity fields then it will gradually radiate energy and angular momentum through its coupling to gravitational radiation (and other massless fields). Hence it will gradually lose energy and its trajectory will approach a geodesic which minimizes the energy. But these trajectories are precisely the zero-energy null geodesics tangent to $V$ on the evanescent ergosurface. Hence the trajectory of our particle will approach one of these trapped null geodesics. It will have very small energy as measured at infinity. However, since the massive particle is now following an almost null trajectory, the energy measured by a {\it local} observer will be enormous. Hence its backreaction on the geometry will be large. This strongly suggests an instability. 

What would be the endpoint of such an instability? The instability involves removing angular momentum from the microstate geometry via radiation. This will tend to shrink the evanescent ergosurface. An obvious candidate endpoint is an almost supersymmetric black hole with the same conserved charges as the microstate geometry, but different angular momenta. This could be a near-extremal BMPV black hole \cite{Breckenridge:1996is} or black ring \cite{Elvang:2004rt}. 

This heuristic argument for instability involves a massive particle. Is there also an instability involving only massless supergravity fields? Our argument relied on the fact that the particle can radiate, i.e., interactions are important. This suggests that a corresponding instability in supergravity will be a nonlinear effect, which makes demonstrating its existence difficult. But it is easy to see why the presence of an evanescent ergosurface makes nonlinear stability unlikely, as we will now explain. 

Proofs of nonlinear stability, e.g., the stability of Minkowski spacetime \cite{Christodoulou:1993uv}, involve first establishing that solutions of the {\it linearized} problem decay sufficiently rapidly. This decay occurs via dispersion to infinity (or across a black hole horizon). Without sufficiently fast decay in the linearized problem there is no reason to expect stability in the nonlinear problem. For example, in anti-de Sitter spacetime (AdS), linear perturbations do not decay. This led to the conjecture that AdS suffers from a nonlinear instability \cite{dafermos}. Such an instability was subsequently discovered numerically \cite{Bizon:2011gg}. 

Supersymmetric microstate geometries are asymptotically flat, so it is possible for linear perturbations to disperse to infinity. However, the presence of the evanescent ergosurface implies that generic linear perturbations decay very slowly because of the stable trapping. To discuss this in more detail, we note first that there exist decoupled linear perturbations that behave like a massless scalar in these geometries \cite{Cardoso:2007ws}. Therefore we will consider the behaviour of a massless scalar field, i.e., the wave equation. Using geometric optics, one can construct low energy, spatially localized, solutions of the wave equation describing wavepackets propagating along the zero energy null geodesics \cite{Sbierski:2013mva}. These can decay by dispersion to infinity but, because of the stable trapping, this involves tunnelling through a potential barrier and so the decay will be very slow. This has been studied in detail for other examples of spacetimes with stable trapping, namely anti-de Sitter black holes \cite{Holzegel:2013kna} and "ultracompact" neutron stars (stars with a photon sphere) \cite{Keir:2014oka}. In both cases, it has been shown that the stable trapping implies that the late time decay is generically as an inverse power of $\log t$ where $t$ labels a foliation by spacelike surfaces such that $\partial/\partial t$ is Killing. (This can be contrasted with the power-law decay of waves in asymptotically flat black hole spacetimes.)

This slow decay presents a serious problem for attempts to prove stability for a {\it nonlinear} equation. Even $t^{-1}$ decay, (as for the linear wave equation in 4d Minkowski space) is problematic, and will generically lead to solutions which blow up in finite time \cite{John:1981}, unless certain conditions are placed on the nonlinearities. An example of such a condition is the "null condition" \cite{Klainerman:1984}. Physically, this condition prohibits interactions between wave packets which are travelling in the same null direction, so, although these waves may remain close to each other for a long time, they cannot interact in order to produce a singularity. It is sometimes possible to replace this condition with a weaker one (the ``weak null condition'' \cite{Lindblad:2004ue}) and still obtain global solutions \cite{Lindblad:2008}. Indeed one can prove the nonlinear stability of Minkowski spacetime this way \cite{Lindblad:2004ue}. 

Given the difficulties already encountered when linear waves decay at a rate $t^{-1}$, slower rates appear particularly troubling. In the case where these rates are related to the phenomenon of stable trapping, the physical mechanism underlying the null condition also appears to be absent: waves can be localised along \emph{different} null directions, but still interact for a long time. This appears particularly dangerous in the case where the stable trapping is ``local'', i.e., confined to a finite region of space, as in microstate geometries and ultracompact stars.\footnote{It is conceivable that the stable trapping may be less of a problem for the example of AdS black holes because there the trapping occurs at infinity. Ref. \cite{Dias:2012tq} argues that such spacetimes will be nonlinearly stable.}

For supersymmetric microstate geometries, the stable trapping appears worse than the other two examples just discussed because the associated null geodesics have zero energy. For the wave equation, the corresponding statement is that the energy degenerates on the evanescent ergosurface, so that smallness of the energy does not imply smallness of the gradient of the field there. This means that standard methods for establishing boundedness of solutions of the wave equation do not work. So even proving {\it linear} stability of the wave equation in these geometries is non-trivial. Even if linear stability can be established, we expect the decay of linear perturbations will be at least as slow as the examples of stable trapping just mentioned, which is far too slow for establishing nonlinear stability.

In the discussion so far we have concentrated on microstate geometries from the 5d perspective. However, such geometries are often best viewed as solutions in 6 dimensions, with a compact Kaluza-Klein circle (indeed some geometries are smooth in 6d but not in 5d). We explain below how to define the evanescent ergosurface from the 6d perspective. We will also investigate the trapping in 6d. Surprisingly, we find that for any supersymmetric microstate geometry, there is a stably trapped null geodesic passing through {\it every} point of the 6d spacetime, i.e., not just points on the evanescent ergosurface. Away from the ergosurface, these correspond to BPS charged particle trajectories in 5d. In this paper, we will focus mainly on the stable trapping on the evanescent ergosurface.

To gain some understanding of the behaviour of geodesics and linear perturbations of microstate geometries, we will study in detail two classes of solutions. In section \ref{sec:3charge} we study the 3-charge microstate geometries of Ref. \cite{Giusto:2004id}. In Appendix \ref{app:2charge} we study the maximally rotating 2-charge microstate geometries of Ref. \cite{maldacena2000}. These solutions are special because they have extra symmetries which enable the geodesic equation or wave equation to be separated and reduced to ODEs. We will show that there are families of quasinormal modes which are localized around the stably trapped zero energy null geodesics on ${\cal S}$, and which decay very slowly. We construct these modes using a matched asymptotic expansion valid for large "total angular momentum" quantum number $\ell \gg 1$, with the result that these
modes have frequency
\be
 \omega \approx \omega_R -i \beta e^{- 2 \ell \log \ell} 
\ee
where $\omega_R$ and $\beta>0$ are constants that are independent of $\ell$ to leading order. There are also quasinormal modes localized around the stably trapped null geodesics away from ${\cal S}$, with ${\rm Im} \omega \sim -\exp(-\ell \log \ell)$.
These results are for $\ell \gg 1$ but we have also constructed such quasinormal modes numerically, and find that they decay very slowly even at small $\ell$. 

We can compare this result with the behaviour of quasinormal modes for AdS black holes \cite{Festuccia:2008zx,Gannot:2012pb} or ultracompact stars \cite{Cardoso:2014sna}. There are two important differences. First, in these examples $\omega_R$ is proportional to $\ell$ at large $\ell$ whereas in our case, $\omega_R$ does not scale with $\ell$. This is closely related to the fact that the associated null geodesics have zero energy. Second, for AdS black holes or ultracompact stars, the imaginary part of the frequency of the most slowly decaying quasinormal modes is of the form $e^{-\gamma \ell}$ (for some $\gamma>0$) whereas we have $e^{-2 \ell \log \ell}$. Hence, in our case, the decay of quasinormal modes is slower than in these other examples of stable trapping. We will explain below why this behaviour of the quasinormal modes implies that generic perturbations decay slower than for AdS black holes or ultracompact stars, and therefore cannot exhibit power law decay. A rigorous result proving this slow decay will appear in a companion paper \cite{joepaper}. 

Our construction of the quasinormal modes exploits the special properties of these particular microstate geometries. However, since these modes are localized around the zero energy null geodesics, we expect that the slow decay of these quasinormal modes is a generic feature of spacetimes with an evanescent ergosurface, and hence our conclusion on the slow decay of generic perturbations should apply to any such spacetime. 

Note that the slowest decaying modes are those with the largest angular frequency. This suggests that the nonlinear instability of such geometries will be a short-distance effect, perhaps involving the formation of tiny (uncharged) black holes, as in the AdS instability. Such black holes would then behave as massive particles, accelerate to the speed of light and cause a large backreaction, perhaps triggering collapse of the evanescent ergosurface, with the solution finally settling down to an almost BPS black hole solution with the same conserved charges as the microstate geometry, but different angular momenta.

The "fuzzball proposal" conjectures that supersymmetric microstate geometries provide a geometrical description of certain quantum microstates of supersymmetric black holes \cite{Mathur:2005zp}.  It is therefore interesting to compare whether the decay of linear waves in a microstate geometry resembles the decay for a supersymmetric black hole. For a supersymmetric black hole, waves are expected to decay as an inverse power law of time at late time outside the horizon. This has been proved for the extremal Reissner-Nordstrom spacetime \cite{Aretakis:2011ha,Aretakis:2011hc}. The slowest decaying modes are those with the lowest angular frequency. However, for a microstate geometry, the stable trapping implies that the decay is slower than any inverse power law, and the slowest decaying modes are those with the highest angular frequency.  Hence there is a qualitative differences between the behaviour of linear waves in microstate geometries and in supersymmetric black hole geometries. 

Another family of spacetimes with an evanescent ergosurface are supersymmetric "black lens" solutions \cite{Kunduri:2014kja,Tomizawa:2016kjh}. A black lens is a black hole with an event horizon of lens space topology. These solutions have an evanescent ergosurface outside the event horizon. Other examples of solutions with this property are obtained by superposing black holes with microstate geometries \cite{Kunduri:2014iga}. Our heuristic particle argument for instability may not apply to these solutions because the particle can fall across the horizon. However, the presence of the evanescent ergosurface implies that it is likely that all of these solutions will exhibit slow decay of linear perturbations and a corresponding nonlinear instability.  

To define the evanescent ergosurface we need a Kaluza-Klein symmetry in 6d. It has been argued that there exist microstate geometries without such a symmetry \cite{Giusto:2013rxa}. (See also Ref. \cite{Bena:2016ypk} and references therein.) In such geometries one cannot define an evanescent ergosurface. Nevertheless, as we will explain, we expect such geometries to admit stably trapped null geodesics through every point of the spacetime. Hence we expect that such geometries will suffer from slow decay of linear perturbations and a corresponding nonlinear instability. 

This paper is organized as follows. In section \ref{sec:geodesics} we review the notion of an evanescent ergosurface in 5d and 6d and show that such a surface is ruled by zero energy null geodesics. For supersymmetric microstate geometries we prove that these geodesics exhibit stable trapping. We also show that a 6d microstate geometry has a stably trapped null geodesic through every point of the spacetime. We elaborate on our heuristic argument for why these geometries are unstable. We then explain why the evanescent ergosurface presents a problem for proving linear stability of these geometries. Even if this problem can be overcome, we argue that the methods required will not extend to the nonlinear problem. In section \ref{sec:3charge} we discuss in detail the 3-charge microstate geometries of Ref. \cite{Giusto:2004id}. In section \ref{sec:quasinormal} we determine quasinormal modes of these geometries in two ways: first using a matched asymptotic expansion (valid for large $\ell$), and then using numerical methods (for general $\ell$). We then explain why the properties of these quasinormal modes imply that generic linear perturbations must decay very slowly, in particular they cannot exhibit power-law decay. Appendix \ref{app:2charge} performs quasinormal mode calculations for the 2-charge microstate geometries of Ref. \cite{maldacena2000}. 

\section{Geodesics and stable trapping}

\label{sec:geodesics}

\subsection{Zero energy null geodesics}

Supersymmetric solutions of 5d supergravity admit a non-spacelike Killing vector field $V$ which approaches a standard time translation at infinity. In a 5d microstate geometry spacetime, $V$ is timelike everywhere except on the evanescent ergosurface: a timelike hypersurface ${\cal S}$, on which $V$ is null. In fact supersymmetry implies that there exists a scalar $f$ such that \cite{Gibbons:1993xt}
\be
V^2 =-f^2
\ee
and ${\cal S}$ is given by $f=0$. Since $V$ is Killing, it preserves ${\cal S}$, i.e., $V$ is tangent to ${\cal S}$. It is easy to see that $V$ is tangent to affinely parameterized null geodesics on ${\cal S}$ \cite{Niehoff:2016gbi}:
\be
\label{geodesic}
 V^b \nabla_b V_a = - V^b \nabla_a V_b = -(1/2)\nabla_a (V^2) 
\ee
and the RHS vanishes on ${\cal S}$ because $V^2$ has a second order zero on ${\cal S}$. Hence $V$ is tangent to affinely parameterized null geodesics on ${\cal S}$.\footnote{Note that this is {\it not} true for a general ergosurface (e.g. in the Kerr spacetime), when $V^2$ has only a first order zero and so the RHS is non-zero and orthogonal to ${\cal S}$ hence $V$ is non-geodesic in that case.} The conserved energy of a timelike or null geodesic with momentum $P^a$ is 
\be
E=-V\cdot P \ge 0
\ee
where the inequality follows because $V$ is non-spacelike and $V,P$ are both future-directed. 

Since $V$ is null on ${\cal S}$, it follows that $V$ is tangent to {\it zero energy} null geodesics on ${\cal S}$. Furthermore, these are the only causal curves with zero energy: away from ${\cal S}$, $V$ is timelike so $E=0$ would imply that $P$ is spacelike whereas on ${\cal S}$, $E=0$ implies that $P$ is tangent to $V$.

Microstate geometries carry non-zero angular momentum. Since $V$ approaches a standard time translation at infinity, a particle following an orbit of $V$ does not rotate w.r.t. to infinity, i.e., it has zero angular velocity. This means that the particle is resisting the frame-dragging effect arising from the rotation of the spacetime geometry. In this sense, the zero energy null geodesics can be regarded as having angular momentum opposite in sign to the angular momentum of the background geometry. If the microstate geometry has appropriate rotational symmetries then one can use these to define conserved angular momenta for geodesics; we will see below that at least one of the angular momenta of the zero energy null geodesics has opposite sign to that of the background. 

\subsection{The 6d perspective}

\label{subsec:6d}

Sometimes it is more convenient to discuss microstate geometries in 6d rather than 5d. In particular, this is the case for 2-charge microstate geometries, and the 3-charge geometries of Ref. \cite{Giusto:2004id}, which are regular in 6d but not in 5d. Therefore we will need to discuss how ${\cal S}$ is defined in 6d. 

The 5d Killing field $V$ is the Kaluza-Klein reduction of a 6d Killing field, which we will also call $V$. Supersymmetry implies that $V$ is globally null w.r.t. the 6d metric \cite{Gutowski:2003rg}. It can be written as $V=T+Z$ where $T$ and $Z$ are commuting Killing vector fields, $Z$ is the spacelike "Kaluza-Klein" Killing vector field (i.e. the 5d metric is obtained from the 6d metric by projecting orthogonally to $Z$ and rescaling) and, near infinity, $T$ is timelike and canonically normalized. 

$V$ is null in 5d if, and only if, it is orthogonal to $Z$ in 6d. Hence, in 6d, ${\cal S}$ can be defined as the locus where $V$ is orthogonal to $Z$. On ${\cal S}$ we therefore have (using the fact that $V$ is null)
\be
 T^2  = Z^2 = -T \cdot Z
\ee
For 2-charge microstate geometries, which do not correspond to regular 5d solutions, ${\cal S}$ is a 2d timelike submanifold on which $Z$ vanishes (and hence $T$ is null). For 3-charge microstate geometries, ${\cal S}$ is a timelike hypersurface in 6d (i.e. codimension 1). In the 3-charge case, $Z$ is non-vanishing on ${\cal S}$ so the above equations imply that $T$ is spacelike on ${\cal S}$. Since $T$ generates time translations in 6d, it follows that there is a genuine ergoregion present in 6d (this has been noticed before \cite{Jejjala:2005yu}). 

In 6d, since $V$ is globally null, it is everywhere tangent to affinely parameterized null geodesics. We use $T$ to define the energy of geodesics in 6d: $E_6=-T\cdot P$ where $P$ is the momentum of the geodesic. We define the Kaluza-Klein momentum as $p = Z\cdot P$. We can use $-V \cdot P \ge 0$ to obtain\footnote{In the 2-charge microstate geometries, $V' = T-Z$ is also a globally null Killing vector field, which implies $E_6 \ge |p|$.} $E_6 \ge p$. Hence the 6d energy is bounded below despite the presence of the ergoregion. Since $V \cdot Z=0$ on ${\cal S}$ it follows that the null geodesics on ${\cal S}$ with tangent $V$ have zero Kaluza-Klein momentum $p=0$ as well as zero 6d energy $E_6=0$. 

\subsection{Stable trapping}

\label{subsec:trapping}

A geodesic is trapped if it "remains within a bounded region of space". Clearly this is true for the zero energy null geodesics on ${\cal S}$ discussed above. We will now show that the null geodesics on ${\cal S}$ tangent to $V$ are {\it stably} trapped in the sense that initially nearby null geodesics remain nearby. This is intuitively obvious since these geodesics minimize the energy; we will now see it explicitly using the geodesic deviation equation (i.e. Jacobi fields). We will also show that {\it all} null geodesics with tangent $V$ are stably trapped in a 6d supersymmetric microstate geometry, hence there exists a stably trapped null geodesic through every point of the spacetime. 

We will first consider a more general situation of $d$-dimensional spacetime admitting a Killing vector field $V$. We define ${\cal T}$ to be the locus where $V^2$ is extremized, i.e., where $\nabla_a (V^2)=0$. Using Killing's equation as in (\ref{geodesic}) we then have $V^b \nabla_b V^a=0$ on ${\cal T}$. Since $V$ must be tangent to ${\cal T}$, we have a family of affinely parameterized geodesics on ${\cal T}$ with tangent $V$. 

Let $\gamma$ denote one of the geodesics on ${\cal T}$ with tangent $V$.  Consider a 1-parameter family of affinely parameterized geodesics which contains $\gamma$ \cite{wald}. Let $X^a$ denote the tangent vector to these geodesics, and $Y^a$ a deviation vector within this family, i.e., ${\cal L}_X Y=0$. On $\gamma$ we have $X^a=V^a$. We want to determine how $Y^a$ behaves along $\gamma$. The geodesic deviation equation gives
\be
 \left(\nabla_V \nabla_V  Y^a \right)|_\gamma =  \left(\nabla_X \nabla_X \right) Y^a |_\gamma= R^a{}_{bcd} X^b X^c Y^d |_\gamma = R^a{}_{bcd} V^b V^c Y^d |_\gamma
\ee
To evaluate the RHS we used the Killing vector identity 
\be
\label{killingcurvature}
 \nabla_c \nabla_a V_b = R_{bacd} V^d
\ee
This implies
\be
 R_{bacd} V^a V^d = \nabla_c \left( V^a \nabla_a V_b \right) - \left( \nabla_c V^a \nabla_a V_b \right) = H_{bc} + \omega^a{}_c \omega_{ab} 
\ee
where
\be
 H_{ab} =H_{ba}= \nabla_a \nabla_b (-V^2/2)
\ee
and
\be
 \omega_{ab} = -\omega_{ba}= \nabla_b V_a
\ee
The geodesic deviation equation is therefore
\be
\label{geodev}
 \left[ \nabla_V \nabla_V  Y^a + \left( H^a{}_b + \omega^{ca} \omega_{cb} \right)Y^b \right]_\gamma=0
\ee
It will be convenient to rewrite this in terms of the Lie derivative w.r.t. $V$ as follows: 
\be
\label{Lie2V}
 {\cal L}_V {\cal L}_V Y^a = \nabla_V \nabla_V Y^a - (\nabla_V Y^b) \nabla_b V^a - Y^b V^c \nabla_c \nabla_b V^a - ({\cal L}_V Y^b) \nabla_b V^a
\ee 
The identity (\ref{killingcurvature}) implies that the 3rd term on the RHS of (\ref{Lie2V}) is zero. The first term is given by (\ref{geodev}). Using this, (\ref{Lie2V}) becomes
\be
  \left( {\cal L}_V {\cal L}_V Y^a  +2 \omega^a{}_b {\cal L}_V Y^b + H_{ab} Y^b \right)_\gamma = 0
\ee
This is a second order ODE governing the evolution of $Y^a$ along $\gamma$. Note that
\be
{\cal L}_V \omega_{ab} = {\cal L}_V H_{ab} = 0
\ee 
which implies that (\ref{Lie2V}) admits the first integral
\be
\label{geocons}
 \left({\cal L}_V Y_a \right) \left( {\cal L}_V Y^a \right) + H_{ab} Y^a Y^b = C
\ee
where $C$ is constant along the geodesic. 

Now we assume that $\gamma$ is a {\it null} geodesic and that $Y^a$ is a deviation vector pointing to a nearby causal geodesic. To do this we consider a 1-parameter family of causal geodesics, so $X^2 \le 0$. Since $X^2=0$ on $\gamma$, we see that $X^2$ is maximized on $\gamma$ within our 1-parameter family. Hence on $\gamma$ we have
\be
 0 = \nabla_Y (X^2) = 2 X^b Y^a \nabla_a X_b = 2 X^b X^a \nabla_a Y_b = 2 X^a \nabla_a (X \cdot Y)
\ee
where we used ${\cal L}_X Y=0$ and the geodesic equation for $X$. It follows that $X \cdot Y$ is constant along $\gamma$, therefore $V \cdot Y$ is constant along $\gamma$ so $V_a {\cal L}_V Y^a = 0$. Hence ${\cal L}_V Y^a$ must be spacelike or null so the first term in (\ref{geocons}) is non-negative. 

Note that $H_{ab}$ is the Hessian of $-V^2/2$, which is extremized on ${\cal T}$. Therefore $H_{ab}$ has components only in directions normal to ${\cal T}$. If assume that ${\cal T}$ is a timelike submanifold then these normal directions are all spacelike. If $-V^2/2$ is {\it minimized} on ${\cal T}$ (as for a microstate geometry) then $H_{ab}$ will be positive semi-definite, so we deduce that $C \ge 0$. Generically, $H_{ab}$ will be positive definite when restricted to the space of vectors normal to ${\cal T}$. In this case, $H_{ab}$ is a Riemannian metric on the space of vectors normal to ${\cal T}$. But we know that $H_{ab} Y^a Y^b \le C$ hence the components of $Y^a$ normal to ${\cal T}$ remain bounded. In other words, at the (infinitesimal) level of geodesic deviation, causal geodesics near to $\gamma$ cannot move away from ${\cal T}$. 

For a 5d supersymmetric microstate geometry, ${\cal T}$ coincides with the evanescent ergosurface ${\cal S}$, which is a hypersurface (i.e. a 4d submanifold). Furthermore, $V^2$ has a {\it second order} zero on ${\cal S}$. This implies that the Hessian can be written $H_{ab} = \alpha n_a n_b$ where $\alpha>0$ is constant along $\gamma$ and $n_a$ is a unit spacelike normal to ${\cal S}$. The argument of the previous paragraph then gives $(n \cdot Y)^2 \le C/\alpha$ hence the component of $Y$ normal to ${\cal S}$ remains bounded so we have stable trapping in the direction normal to ${\cal S}$. Hence causal geodesics that are initially close to $\gamma$ will remain close to ${\cal S}$. 

Now consider the case in which $V$ is globally null, e.g. a supersymmetric microstate geometry in 6d. In this case ${\cal T}$ is the entire spacetime and $H_{ab}$ vanishes. However, we can see stable trapping as follows. From (\ref{killingcurvature}) we see that $\nabla_V \omega_{ab}=0$ so the geodesic deviation equation (\ref{geodev}) admits a first integral\footnote{Note that we cannot do this when $H_{ab} \ne 0$ because $\nabla_V H_{ab} \ne 0$ in general. The constants $C$ and $C'$ differ by a multiple of $\omega_{ab} Y^a {\nabla_V Y^b}$ which can be shown to be constant along $\gamma$ using (\ref{geodev}).}
\be
 (\nabla_V Y_a) (\nabla_V Y^a) + \omega_{ac} \omega_b{}^c Y^a Y^b = C'
\ee
where $C'$ is constant along the geodesic. As above, $V \cdot Y$ is constant along a geodesic $\gamma$ with tangent $V$ so $V_a \nabla_V Y^a=0$. Hence $\nabla_V Y^a$ is spacelike or null so the first term above is non-negative.  Hence we have
\be
\label{omegaYbound}
 \omega_{ac} \omega_b{}^c Y^a Y^b  \le C''
\ee
for some new constant $C''$. Note that the LHS is non-negative because $\omega_{ab}$ is orthogonal to $V$ hence $\omega_{ab} Y^b$ is non-timelike. 

Note that $\omega_{ab}$ is the rotation of the null geodesic congruence with tangent $V$.\footnote{We emphasize that our 1-parameter family is not assumed to belong to this congruence, i.e., $Y^a$ is a general deviation vector, not necessarily one associated with this congruence.} As is usual when dealing with such a congruence, we can pick a null basis $\{ e_\mu^a \}$ where $e_0=V$ and $e_1$ is null with $e_0 \cdot e_1 = -1$ and $e_i$ $(i=2,3,\ldots, d-1)$ are orthonormal spacelike vectors orthogonal to $e_0$ and $e_1$. Furthermore, we can choose our basis to be parallelly transported along the geodesics of the congruence. In such a basis, the components $\omega_{\mu\nu}$ are constants along $\gamma$ and $\omega_{0\mu}=0$. Equation (\ref{omegaYbound}) becomes
\be
 \left( \omega_{i1} Y^1 + \omega_{ij} Y^j \right)\left(  \omega_{i1} Y^1 + \omega_{ij} Y^j \right) \le C''
\ee
Next note that $Y^1 = -e_0 \cdot Y = - V\cdot Y$, which we showed above is constant along $\gamma$. Hence $\omega_{i1} Y^1$ is constant along $\gamma$ so it follows from this equation that $\omega_{ij} Y^j$ is bounded (w.r.t. the norm $\delta_{ij}$). 

Now assume that our spacetime contains an evanescent ergosurface ${\cal S}$, i.e., a timelike surface with equation $Z \cdot V=0$. Any covector normal to ${\cal S}$ is parallel to
\be
 n_a = \nabla_a ( Z \cdot V ) = Z^b \nabla_a V_b + V^b \nabla_a Z_b = -Z^b \nabla_b V_a - V^b \nabla_b Z_a = -2 Z^b \nabla_b V_a = -2 \omega_{ab} Z^b
\ee 
with $n_a$ spacelike (because ${\cal S}$ is timelike). Note that
\be
 n \cdot Y = 2 \omega_{ab} Z^a Y^b = 2 \omega_{ij} Z^i Y^j
\ee
where we used $Z^1 = Z \cdot V=0$. We have just shown that $\omega_{ij} Y^j$ is bounded along $\gamma$, hence $n \cdot Y$ is also bounded. It follows that ${\cal S}$ exhibits stable trapping: deviation vectors cannot become large in the direction orthogonal to an evanescent ergosurface ${\cal S}$. 

We can deduce a little more from the above analysis. We no longer assume that $\gamma$ is on ${\cal S}$. We showed above that, along $\gamma$, $Y^1$ is constant and $\omega_{ij} Y^j$ is bounded. Now {\it assume} that $\omega_{ij}$ is non-degenerate. It follows that $Y^j$ must be bounded along $\gamma$. In fact, it is easy to solve explicitly the geodesic deviation equation (\ref{geodev}) to see that $Y^i$ oscillates along $\gamma$, such that the mean value of $\omega_{ij} Y^j$ is $-\omega_{i1} Y^1$. One can then solve for $Y^0$, finding an oscillating term plus a term that grows linearly. The latter is "pure gauge": it can be eliminated by a change of affine parameter along the geodesics of the 1-parameter family. Having done this, all components of $Y^a$ are bounded along $\gamma$. This is stable trapping. Hence if the congruence of null geodesics with tangent $V$ has non-degenerate rotation matrix $\omega_{ij}$ then {\it any} geodesic in this congruence exhibits stable trapping. The constant $Y^1$ represents a shift from a geodesic $\gamma$ in this congruence to a nearby geodesic $\gamma'$ also within this congruence and the deviation vector describes oscillations about $\gamma'$. 

We can apply this argument to supersymmetric microstate geometries in 6d.\footnote{
In 10d, $\omega_{ij}$ is degnerate in directions associated with the internal $T^4$. However, the compactness of this space prevents the geodesics from dispersing in these directions.} We will show later that $\omega_{ij}$ is indeed everywhere non-degenerate for the 3-charge microstate geometries of \cite{Giusto:2004id,Giusto:2004ip,Giusto:2004kj}, and also the 2-charge geometries of \cite{maldacena2000,balasubramanian2000}. It seems very unlikely that more complicated microstate geometries would have degenerate $\omega_{ij}$ so we expect $\omega_{ij}$ to be non-degenerate for general supersymmetric microstate geometries (including those lacking the Kaluza-Klein Killing vector field $Z$ as in \cite{Giusto:2013rxa}). So we expect that the null geodesics with tangent $V$ are all stably trapped in any supersymmetric microstate geometry. Hence there is a stably trapped null geodesic through every point of the 6d spacetime. Of course, these include the zero energy null geodesics on ${\cal S}$, which are singled out by the additional condition of having zero Kaluza-Klein momentum. 

Away from ${\cal S}$ the stably trapped null geodesics have non-zero Kaluza-Klein charge $p$. From the 5d perspective, these null geodesics look like "BPS" charged particles, i.e., with mass equal to charge, which are at rest relative to infinity. It is familiar that such particles can remain at rest because they experience a cancellation of forces. But often this corresponds to neutral equilibrium (degenerate $\omega_{ij}$, which allows linear growth of deviation vectors), whereas we have stable equilibrium. It would be interesting to investigate how this stability arises from the interaction of the particle with the various 5d fields. 

In arguing for instability, we will focus on the consequences of the stable trapping on ${\cal S}$ because in this case we have stable trapping of null geodesics in 5d as well as is 6d. The consequences of the stable trapping away from ${\cal S}$ in 6d would be interesting to explore further. 

\subsection{Heuristic argument for instability}

\label{sec:heuristic}

In the Introduction, we presented a heuristic argument that supersymmetric microstate geometries experience an instability because a massive uncharged 5d particle will accelerate to the speed of light on ${\cal S}$, and cause a large backreaction. We will now discuss this in more detail. 

Let $\Sigma_0$ be a spacelike Cauchy surface for a  5d microstate geometry. Choose coordinates $x^i$ on $\Sigma_0$ and let $t$ be the parameter distance from $\Sigma_0$ along the integral curves of $V$. Carry the coordinates $x^i$ along these integral curves to define coordinates $(t,x^i)$. The metric can then be written in ADM form
\be
 ds^2 = -N^2 dt^2 + h_{ij} \left( dx^i - \Omega^i dt \right)\left( dx^j - \Omega^j dt \right) 
\ee
where 
\be
 N^2 = f^2 + h_{ij} \Omega^i \Omega^j,
\ee
$V = \partial/\partial t$ is the stationary Killing vector field, and $f=0$ on ${\cal S}$. In general there is freedom to shift $t$ by a function of the other coordinates.

For the 3-charge microstate geometries that we will study later,\footnote{These have a pair of orbifold singularities when reduced to 5d but that is not relevant to this argument.} we can split the coordinates as $x^i=(x^I,x^\alpha)$ such that $\partial/\partial x^I$ ($I=1,2$) are Killing vectors associated to rotational symmetries, and $\Omega^\alpha=0$, and it is natural to chose $\Sigma_0$ so that $\partial/\partial x^I$ are tangent to it, which eliminates the freedom to shift $t$. 

We will consider a family of local observers whose velocity is othogonal to surfaces of constant $t$. The velocity of such an observer is
\be
 u^a = -N (dt)^a = \frac{1}{N} \left( \frac{\partial}{\partial t} + \Omega^i \frac{\partial}{\partial x^i} \right)
\ee
For a microstate geometry with rotational symmetries, the velocity of these observers is orthogonal to $\partial/\partial x^I$ and so they have zero angular momentum. Hence they are referred to as "zero angular momentum observers" (ZAMOs). Note that they rotate with angular velocities $\Omega^I$ w.r.t. a stationary observer at infinity. This is because of the frame-dragging caused by the rotation of the spacetime. For a general microstate geometry we don't expected any rotational symmetries but we will still refer to these observers as ZAMOs. In general there is the freedom to shift $t$ by a function of $x^i$ so there are many different families of ZAMOs. 

Now consider a particle with mass $\mu$. Its momentum $P_a$ obeys
\be
 -\mu^2 = g^{ab} P_a P_b
\ee
which can be rearranged to give
\be
\label{EJ}
 E^2 - 2 EJ- \frac{f^2}{h_{jk} \Omega^j \Omega^k} J^2 = N^2 \left( \mu^2 + H^{kl} P_k P_l \right) \equiv \Delta^2
\ee
Here $E=-P_t \ge 0$ is the energy of the particle (conserved if it follows a geodesic) and
\be
J = \Omega^i P_i
\ee
We have decomposed $P^i$ so that the component of $P_i$ along $\Omega^i$ appears on the LHS of \eqref{EJ} and the orthogonal component appears on the RHS where we have defined $H^{ij}$ to be the projection of $h^{ij}$ orthogonal to $\Omega^i$:
\be
 H^{ij} = h^{ij} - \frac{\Omega^i \Omega^j}{h_{kl} \Omega^k \Omega^l}
\ee
For a microstate geometry with rotational symmetries, we have $J=\Omega^I P_I$ and $P_I$ are the angular momenta of the particle, which are conserved if the particle follows a geodesic. 

Note that the energy of the particle according to a ZAMO is
\be
 E_{\rm ZAMO} = -u \cdot P = \frac{1}{N} ( E - J )
\ee

\begin{figure}[h]
\centering
\includegraphics[width=0.4\linewidth]{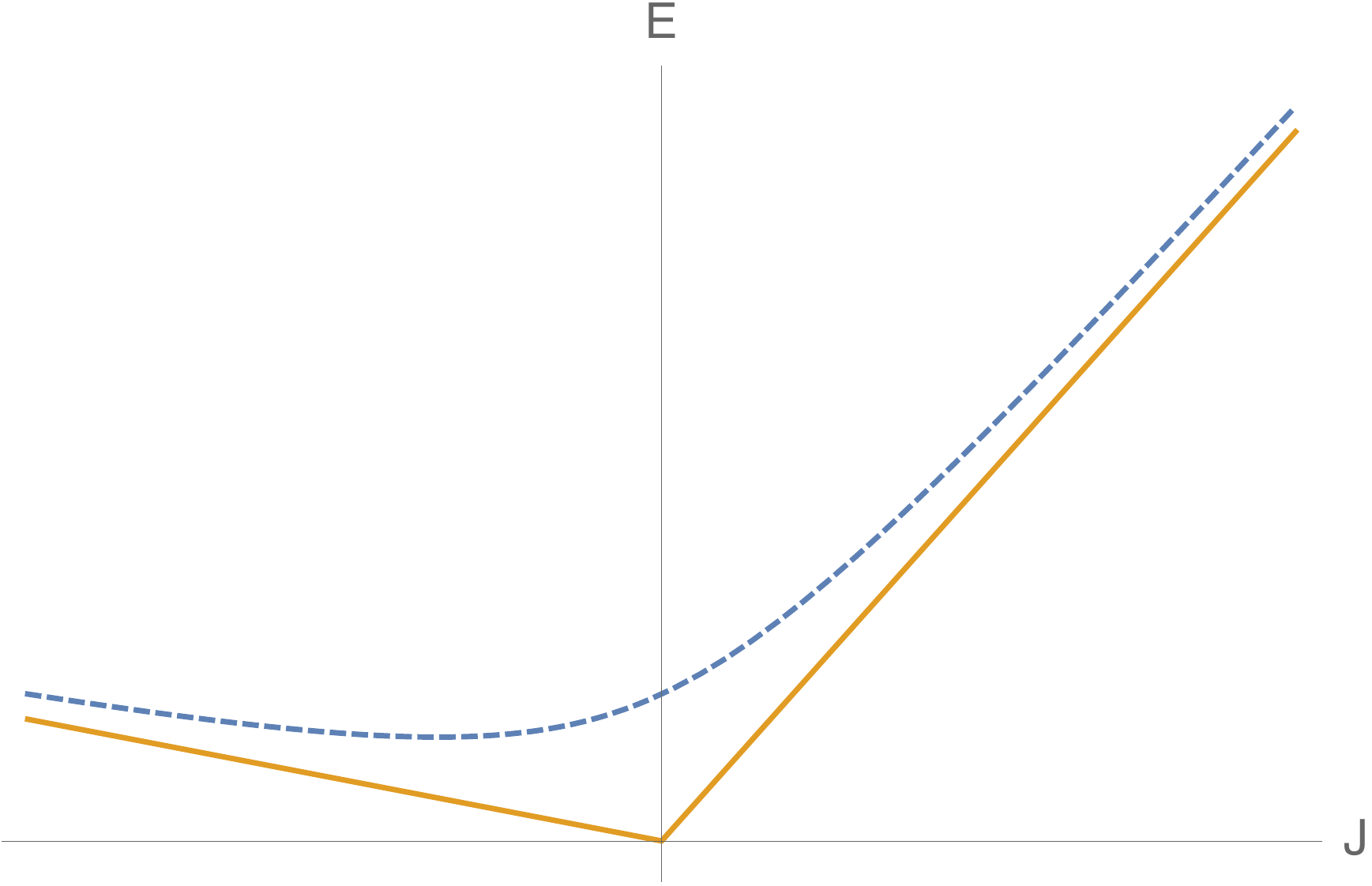}\hspace{1.5cm}
\includegraphics[width=0.4\linewidth]{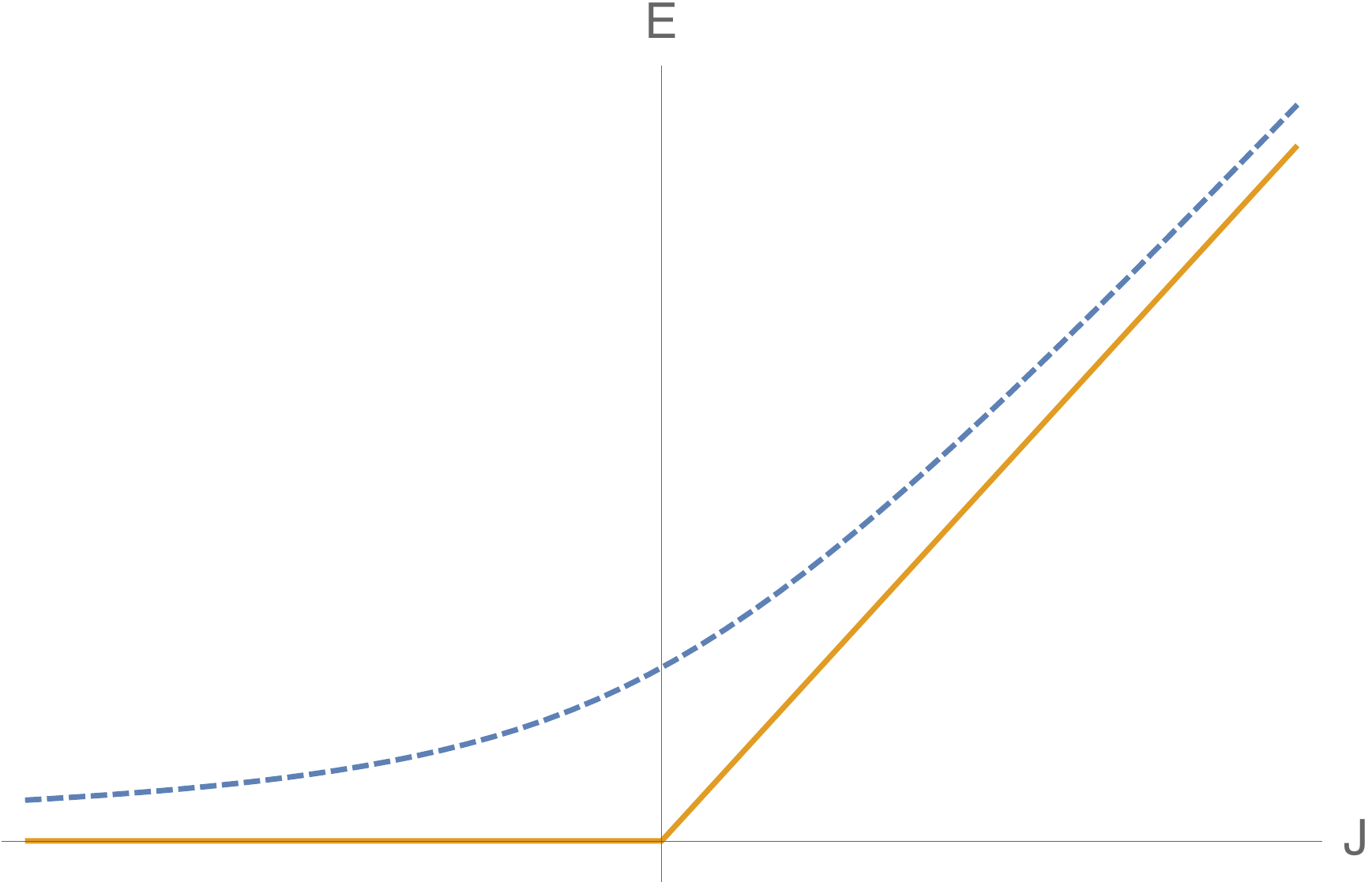}
\caption{Plots of $E$ against $J$. Dashed blue curves: $\Delta >0$, solid orange lines: $\Delta=0$. Left: a generic point of spacetime. Right: on an evanescent ergosurface.}
\label{EvsJ}
\end{figure}
To formulate our argument for instability, it is useful to consider equation \eqref{EJ}. At a generic point of a microstate geometry spacetime we have $f \ne 0$ and Figure \ref{EvsJ} (left) shows $E$ as a function of $J$ for fixed $\Delta$. The minimum value of $E$ is positive and occurs at some finite value of $J$. However, at an evanescent ergosurface, we have $f=0$ and the corresponding figure is shown on the right of Fig. \ref{EvsJ}. If $\Delta>0$ then $E$ is minimized at $J= -\infty$. 

First consider a massive particle $\mu>0$. If the particle is free then it will move on a geodesic, so $E$ is conserved. However, when interactions are included, the particle couples to gravitational radiation (and other massless field), and therefore gradually loses energy through radiation. If $E<\mu$ then the particle cannot escape to infinity. Its energy $E$ will decrease over time and approach its minimum value. From the plots, it is clear that the energy is minimized on the evanescent ergosurface, and this minimum occurs at $J=-\infty$ for a massive particle (as $\Delta>0$). Hence the particle must "roll down the hill" to $J=-\infty$. This implies that $E_{\rm ZAMO}$ will diverge, i.e., the local observer will measure infinite energy. This strongly suggests that the spacetime will be unstable.\footnote{Note that one could not apply this argument in a supersymmetric black hole spacetime because the particle would fall across the horizon with non-zero $E$.}

Now consider a massless particle, $\mu=0$. If the particle starts on a stably trapped geodesic then it cannot escape to infinity. As for the massive particle, $E$ will gradually decrease so we can apply the above argument when $\Delta>0$. However, it is possible that the particle will radiate in such a way that it approaches a final state with $\Delta=0$, in which case it can eventually reach $E=0$ at finite negative $J$. This corresponds to to one of the null geodesics tangent to $V$ on ${\cal S}$. 
However, there is nothing preventing this endpoint from having arbitrarily large $J$, so one might expect generically that this will be the case simply because there is more phase space available at large $J$. This again suggests instability. 

\subsection{The energy functional}

We will now discuss the consequences of the existence of an evanescent ergosurface for linear perturbations of microstate geometries. We will explain how establishing even {\it linear} stability in such backgrounds is problematic, and then discuss the consequences for nonlinear stability. 

Known microstate geometry solutions can be obtained as solutions of 6d supergravity. For these solutions, Ref. \cite{Cardoso:2007ws} showed that one can identify certain decoupled sectors of linear perturbations for which the 6d equation of motion is simply that of a massless, uncharged, scalar field, i.e., the wave equation. If this field does not vary around the Kaluza-Klein circle then it will also satisfy the wave equation in 5d. 

The usual method for establishing that solutions of the wave equation remain bounded in time is based on the existence of a conserved energy functional. Consider a globally hyperbolic spacetime with a causal Killing vector field $V$. A field $\Phi$ satisfying the wave equation has a conserved energy momentum tensor
\be
 T_{ab} = \partial_a \Phi \partial_b \Phi - \frac{1}{2} g_{ab} \left( \partial \Phi \right)^2
\ee
We can define a conserved energy-momentum current for $\Phi$:
\be
 j^a = -T^a{}_b V^b
\ee
Let $\Sigma_0$ be a spacelike Cauchy surface and let $\Sigma_t$ be the image of $\Sigma_0$ by moving parameter distance $t$ along the integral curves of $V$. The energy of $\Phi$ on $\Sigma_t$ is then
\be
 E_t[\Phi] = -\int_{\Sigma_t} \sqrt{h} \;  n \cdot j
 \ee
 where $h$ is the determinant of the induced metric on $\Sigma_t$ and $n$ is the future-directed unit normal to $\Sigma_t$. 
 
Since $T_{ab}$ satisfies the dominant energy condition, $j^a$ must be causal and future-directed, or zero. This implies that $E_t \ge 0$. Since $j$ is conserved, it follows that if $t'>t$ then we have $E_{t'} \le E_t$. (Here we allow for the possibility of the surfaces extending to future null infinity, in which case energy can be lost by radiation through null infinity.) Hence if $E_0$ is small then $E_t$ remains small for all $t>0$. 

Consider the integrand of $E_t$. The dominant energy condition implies that $-n \cdot j \ge 0$ with equality if, and only if, $j=0$. But $j=0$ implies (by contracting with $d\Phi$) that $V \cdot \partial \Phi=0$ and $(\partial \Phi)^2=0$. If $V$ is timelike then this implies $d\Phi=0$. However, if $V$ is null then it implies only that $d\Phi$ parallel to $V$.

If $V$ is timelike everywhere then $E$ is a positive-definite functional of $d\Phi$, i.e., $E$ defines a norm for $d\Phi$. If there exist additional Killing vector fields $K^I$ that span the tangent space of $\Sigma_t$ then one can commute the wave equation several times with these vector fields to obtain bounds on $E[K^{I_1} \ldots K^{I_N} \Phi]$ and hence control the norm of higher derivatives of $\Phi$. The Sobolev embedding theorem can then be used to bound $\Phi$. This process may be adapted in several ways: the commuting vector fields need not be exactly Killing, they may only span a submanifold of $\Sigma_t$ (e.g.\ \cite{Dafermos:2009uq}), or the commutation may be with higher order, tensorial operators rather than vector fields (e.g.\ \cite{Dafermos:2013bua}). 

Now consider a 5d supersymmetric microstate geometry. In this case, $V$ is null on ${\cal S}$. Hence on ${\cal S}$, $E$ fails to control the component of $d\Phi$ in the direction of $V$ so $E$ is not positive definite and the above argument for demonstrating boundedness of $\Phi$ does not work. Conservation of energy does not prevent $d\Phi$ from becoming large on ${\cal S}$.\footnote{From the 6d perspective, the functional $E$ gives the difference  $E_6 - p$ where $E_6$ is the 6d energy (defined using the Killing field $T$) and $p$ the Kaluza-Klein momentum (defined using the Killing field $Z$). If we restrict attention to fields $\Phi$ invariant around the KK circle, i.e., $Z \cdot \partial \Phi=0$, then we have $p=0$ so $E_6=E \ge 0$. Since $V$ is globally null, $E$ fails everywhere to control the component of $d\Phi$ along $V$. But we have imposed the additional condition $Z \cdot \partial \Phi=0$, so $d\Phi$ can be proportional to $V$ only when $V$ is orthogonal to $Z$, i.e., on ${\cal S}$.}

This problem arises also for stationary black hole geometries, where $V$ becomes null at the horizon. For a non-extremal black hole, this problem is overcome by exploiting the "horizon redshift effect". This arises from the fact that affinely parameterized horizon generators have tangent $e^{-\kappa t} V$ where $\kappa $ is the surface gravity and $t$ is a parameter along the integral curves of $V$. Hence a photon travelling along a horizon generator suffers a redshift $e^{-\kappa t}$. The wave analogue of this effect enables one to control the behaviour of the problematic component of $d\Phi$ at the horizon \cite{Dafermos:2005eh,Dafermos:2008en}. However, this effect is absent for an extremal black hole. In the extremal case, it turns out that the problematic component of $d\Phi$ remains bounded but higher derivatives blow up along the horizon, i.e., there is an instability \cite{Aretakis:2011ha,Aretakis:2011hc,Aretakis:2012ei,Lucietti:2012sf}.

For a supersymmetric microstate geometry, $V$ is tangent to affinely parameterized geodesics on ${\cal S}$ so there is no analogue of the horizon redshift effect that can be used to control the behaviour of $d\Phi$ on ${\cal S}$. To control the problematic component of $d\Phi$ on ${\cal S}$ one might attempt to proceed as follows. First introduce an everywhere timelike vector field $W$ which agrees with $V$ everywhere except near ${\cal S}$. Now use $W$ to define an energy functional. This new energy functional will be non-degenerate (i.e. it defines a norm on $d\Phi$) but non-conserved. The idea is that we can control the problematic component of $d\Phi$ by commuting the wave equation with Killing vector fields or higher order operators. In particular, if the microstate geometry admits angular momentum operators which commute with the wave operator, then we can first commute with these operators, in order to obtain a bound on the associated higher order energy. We could then integrate this bound in time to show that the non-degenerate energy can grow at most linearly in time. But of course this does not exclude an instability. Alternatively, if a version of Hardy's inequality (see e.g. \cite{Dafermos:2010hb}) can be proved on these backgrounds, then a similar argument could be employed in order to show that the nondegenerate energy is bounded for all time. 

These arguments will only work when the background has appropriate symmetries, which will not be the case for a general microstate geometry. Furthermore, even when the background has such symmetries, these arguments are unlikely to extend to the nonlinear problem. In the nonlinear problem we would no longer have an exactly conserved energy so if we were to try to bound the energy of a perturbation by its initial value then we would encounter various error terms. In order to prove stability, we need to bound these error terms in a suitable way in terms of the initial data. This is often done in the context of a bootstrap argument: the error terms are assumed to satisfy certain bounds, which allows the energy to be bounded, and this in turn allows the initial assumptions on the error terms to be verified and improved. However, if we take the approach suggested above for the linear problem, and first commute the equation with (approximate) angular momentum operators, then the error terms will involve higher derivatives of the field, so we will need to assume bounds on higher-order energies in order to be able to bound lower-order energies. However, this scheme can never ``close'' -- in order to bound these higher-order energies, we would need to assume bounds on even higher order energies, and so on.

In summary, the existence of an evanescent ergosurface implies that standard methods for establishing boundedness of solutions of the linear wave equation do not work in supersymmetric microstate geometries. It is conceivable that this problem could be overcome for microstate geometries admitting suitable rotational symmetries. But such geometries are not typical and furthermore, the methods required are not robust enough to extend to the nonlinear problem.

\section{3-charge microstate geometries}

\label{sec:3charge}

\subsection{Metric and charges}

In this section we will study in detail the 3-charge microstate geometries of Refs. \cite{Giusto:2004id,Giusto:2004ip,Giusto:2004kj}. These are supersymmetric solutions of type IIB supergravity compactified on $T^4$. The resulting 6d geometry asymptotically approaches the product of 5d Minkowski spacetime with a Kaluza-Klein circle of radius $R_z$. We will focus on the case for which the 6d geometries are smooth with no conical or orbifold singularities. These geometries can be reduced to 5d however the 5d metric has a pair of orbifold singularities so it is more convenient to work in 6d. 

These solutions admit 4 Killing vector fields and a "hidden" symmetry (associated to a Killing tensor field) which enables one to separate the wave equation (and Hamilton-Jacobi equation for geodesics) into ODEs. 

The 3 charges of these solutions arise from $n_1$ D1-branes wrapped around the Kaluza-Klein $S^1$, $n_2$ D5-branes wrapped around $S^1\times T^4$, and $n_p$ units of momentum around the $S^1$ where
\be
 n_p = n(n+1) n_1 n_2\qquad n \in \mathbb{Z}
\ee
The solution is written in terms of dimensionful charges
\begin{equation}
\Qo=\frac{(2\pi)^4g\alpha'^3}{V}n_1 \qquad \Qf=g\alpha'n_2 \qquad Q_p=a^2n(n+1)=\frac{4G^{(5)}}{\pi R_z}n_p
\end{equation}
where $g$ is the string coupling constant, $V$ is the volume of the $T^4$, $G^{(5)}$ is the 5d Newton constant and the length scale $a$ is defined by
\be
a =\frac{\sqrt{Q_1Q_2}}{R_z} \label{eq:a}
\ee 
The 10d string frame metric is:
 \begin{equation} \begin{split}
ds^2=&-\frac{1}{h}(dt^2-dz^2)+\frac{Q_p}{hf}(dt-dz)^2+hf\Big(\frac{dr^2}{r^2+(\gto+\gtt)^2\eta}+d\theta^2\Big) \\ &+h\Big(r^2+\gto(\gto+\gtt)\eta-\frac{(\gto^2-\gtt^2)\eta Q_1Q_2\cos^2\theta}{h^2f^2}\Big)\cos^2\theta d\psi^2\\ &+h\Big(r^2+\gtt(\gto+\gtt)\eta+\frac{(\gto^2-\gtt^2)\eta Q_1Q_2\sin^2\theta}{h^2f^2}\Big)\sin^2\theta d\phi^2 \\ &+\frac{Q_p(\gto+\gtt)^2\eta^2}{hf}(\cos^2\theta d \psi+\sin^2\theta d\phi)^2 \\&-2\frac{\sqrt{Q_1Q_2}}{hf}\Big(\gto\cos^2\theta d\psi+\gtt \sin^2\theta d\phi\Big)(dt-dz)\\&-2\frac{(\gto+\gtt)\eta\sqrt{Q_1Q_2}}{hf}\Big(\cos^2\theta d\psi+\sin^2 \theta d\phi \Big)dz+\sqrt{\frac{H_1}{H_2}}\Sigma_{i=1}^{4}dx_i^2
\\=& ds^2_6+\sqrt{\frac{H_1}{H_2}}\Sigma_{i=1}^{4}dx_i^2 \end{split} \label{eq:metric} \end{equation}
where 
\begin{equation}
\eta=\frac{Q_1Q_2}{Q_1Q_2+Q_1Q_p+Q_2Q_p}, \label{eq:eta}
\end{equation}
\be
\tilde{\gamma}_1=-an,\;\;\tilde{\gamma}_2 =a(n+1),\
\ee
\bea
f&=r^2+(\gto+\gtt)\eta(\gto\sin^2\theta +\gtt\cos^2\theta) \nonumber \\  &=r^2+a^2\eta(-n\sin^2\theta+(n+1)\cos^2\theta),
\eea
\begin{equation}
H_1=1+\frac{Q_1}{f},\;\;H_2=1+\frac{Q_2}{f}\; \text{ and } \; h=\sqrt{H_1H_2},
\end{equation}
where $\theta\in [0,\pi/2]$, $r>0$ and $0 \le \phi,\psi \le 2\pi$. 

The angular momenta of these geometries are 
\begin{equation}
J_{\psi}=- n n_1n_5 \qquad J_{\phi}=(n+1) n_1n_5,
\end{equation} 
It is worth noting that we will need to work in the Einstein frame in 6d but that when we reduce from 10 to 6 dimensions and then go to the Einstein frame, the factors involved cancel so the 6d Einstein metric is exactly the same as $ds^2_6$, the 6d part of the 10d string frame metric in \eqref{eq:metric}.

\subsection{Evanescent ergosurface and zero energy null geodesics\label{sec:geo}}

The above solution is supersymmetric and therefore admits a globally defined null Killing vector field:
\be
 V = T+ Z 
\ee
where
\be
 T = \frac{\partial}{\partial t} \qquad Z = \frac{\partial}{\partial z}.
\ee
As discussed in section \ref{subsec:6d}, the evanescent ergosurface $\mathcal{S}$ is defined as the surface where the Kaluza-Klein Killing vector field $Z$ is orthogonal to $V$. We have $V\cdot Z=1/h$ and hence ${\cal S}$ is the surface where $h$ diverges, i.e., where $f=0$. Solving the equation $f=0$ for $0<r<\infty$ gives the following ranges of $\theta$ on $\mathcal{S}$ \cite{Giusto:2004ip}: \begin{itemize}
	\item $n>0$: $\theta\in I_{n>0}=[\tilde{\theta},\pi /2] \;\;\text{ where } \tan\tilde{\theta}=\sqrt{\frac{n+1}{n}}$;
	\item $n<0$: $\theta\in I_{n<0}=[0,\tilde{\theta}]. $
\end{itemize}
It was shown in \cite{Giusto:2004kj} that the 6d metric is regular on ${\cal S}$ and that $\mathcal{S}$ has topology $S^1\times S^3$. 

Due to the symmetries of the spacetime, if $U$ is the tangent vector to an affinely parameterized geodesic then the quantities $p_I=(\pd/\pd x^I)\cdot U$ are conserved along the geodesic, where $x^I\in \{t,\,z,\,\phi,\,\psi\}$. As discussed in section \ref{subsec:6d}, $V$ is everywhere tangent to null geodesics. The conserved quantities associated to these geodesics are
\be
p_t = -h^{-1} \qquad p_z = h^{-1} \qquad p_\psi=-\frac{\sqrt{\Qo\Qf}}{hf}a\eta \cos^2\theta \qquad p_\phi = -\frac{\sqrt{\Qo\Qf}}{hf}a\eta \sin^2\theta.
\ee
On ${\cal S}$, these become
\begin{equation}
	p_t=0,\;\;p_z=0,\;\;\ppsi=-a \eta \cos^2\theta,\;\;\pphi=-a\eta \sin^2\theta. \label{eq:Vcons}
\end{equation}
so the energy ($-p_t$) and Kaluza-Klein charge ($p_z$) both vanish on ${\cal S}$, as expected from section \ref{subsec:6d}. 
Note that $p_\phi+p_\psi$ has opposite sign to $J_\phi+J_\psi$; in this sense, the geodesics have angular momenta opposed to those of the background geometry. If we define $J_L = J_\phi-J_\psi$ and $J_R = J_\phi+J_\psi$ then the background geometry has $J_L = (2n+1) n_1 n_5$, $J_R = n_1 n_5$ so if $n,n_1,n_5 \gg 1$ then $J_L \gg J_R \gg 1$. The backreaction of particles following geodesics on ${\cal S}$ will tend to reduce $J_R$ so it is plausible that the final state of the instability will be a near-extremal BMPV black hole \cite{Breckenridge:1996is}, which has $J_R \approx 0$. 

The energy of these geodesics as measured by a {\it local} observer is not small. For example, consider a zero angular momentum observer (ZAMO) (as in section \ref{sec:heuristic}) with velocity $u^a$ given by
\be
 u^a =  -\frac{(dt)^a}{\sqrt{-g^{tt}}} 
  \ee
On ${\cal S}$, a ZAMO measures the energy of a null geodesic with momentum $V$ to be
\be
 E_{ZAMO} = -u \cdot V = \sqrt{\Qo\Qf}\left(\Qo+\Qf+Q_p+\frac{\Qo\Qf+\Qo Q_p+\Qf Q_p}{a^2\eta\left((n+1)\sin^2\theta-n \cos^2\theta \right)}\right)^{-\frac{1}{2}}.
 \ee
As discussed in section \ref{subsec:trapping}, the condition for the null geodesics with tangent $V$ to be stably trapped everywhere is for the rotation matrix $\omega_{ij}$ of the null geodesic congruence with tangent $V$ to be non-degenerate. One can define the rotation as follows \cite{wald}. At any point, consider the space of vectors orthogonal to $V$ quotiented by the subspace of vectors proportional to $V$. This defines a 4d vector space ${\cal V}$, and $\omega = -(1/2)dV$ can be regarded as a 2-form acting on vectors in this space. We want to ask whether this 2-form is non-degenerate. So we need to calculate dV. We start from
\be
 V  = -h^{-1} (dt - dz) +  C (hf)^{-1}  \left (\cos^2 \theta d\psi + \sin^2 \theta d\phi \right) 
\ee
where $C$ is a constant and hence
\begin{equation} \begin{split}
 dV = \frac{1}{2}&\big((\Qo+\Qf)f+2\Qo\Qf\big)(hf)^{-3} \Big[ r(dt-dz)\wedge dr -a^2\eta (2n+1)\sin\theta\cos\theta (dt-dz)\wedge d\theta\Big]\\ 
  & +\frac{C}{2}(2f+\Qo+\Qf) (hf)^{-3}r \Big[\cos^2\theta\, d\psi \wedge dr+\sin^2\theta\, d\phi\wedge dr\Big] \\ 
   &+\frac{C}{2}\sin\theta \cos\theta(hf)^{-3} \Big[ 2 (hf)^2\left(d\psi\wedge d\theta -d\phi\wedge d\theta\right) \\ & \hspace{30mm}-a^2\eta(2n+1)(2f+\Qo+\Qf)\left(\cos^2\theta\,d\psi\wedge d\theta +\sin^2\theta\,d\phi\wedge d\theta\right)\Big].
\end{split} \end{equation}
Now we want to show that this is non-degenerate by acting on an arbitrary vector $X \in {\cal V}$. Since $X \sim X + \alpha V$ we can choose $X$ so that $X^t=0$. The condition $X\cdot V=0$ then fixes $X^z$. We now consider $(dV)_{ab} X^b$ as a covector acting on ${\cal V}$ so we neglect terms proportional to $V_a$ in $(dV)_{ab} X^b$. The result is that this covector vanishes if, and only if, $X^r = X^\theta = X^\phi = X^\psi=0$ and hence $X^z=0$. Therefore $dV$ is non-degenerate, viewed as a quadratic form on ${\cal V}$. Hence the rotation matrix is non-degenerate. By setting $Q_p=0$ one sees that this result applies also to the 2-charge microstate geometries discussed in the Appendix.

\section{Quasinormal modes}	

\label{sec:quasinormal}

\subsection{Relation to null geodesics}

We will now consider the wave equation
 \begin{equation} \Box \Phi =0 \label{eq:wave} \end{equation} 
in the geometry (\ref{eq:metric}). The geometric optics approximation tells us that we can expect to find rapidly varying solutions of this equation which are localized around null geodesics for an arbitrarily long time.\footnote{Furthermore, the results of Ref. \cite{Sbierski:2013mva} prove that the energy of the solution is close to the energy of the corresponding null geodesic.} Therefore we expect there to exist solutions of the wave equation that are localized around a null geodesic with tangent $V$. Of course, such solutions will eventually decay by dispersion to infinity. 

In this section, we will show that such solutions can be constructed as quasinormal modes, i.e., modes with definite frequency $\omega$. For black hole solutions, it is known that quasinormal mode frequencies can be related to properties of trapped null geodesics in the geometric optics limit \cite{Ferrari:1984,Yang:2012he}. For example, consider a Kerr black hole. One can look for mode solutions of the form
\be
 e^{-i\omega t + i m \phi} \Phi_r(r) \Phi_\theta(\theta)
\ee
The angular equation gives spheroidal harmonics labelled by an integer $\ell$ with $|m| \le \ell$. If $\ell \gg 1$ then one can construct families of quasinormal modes with frequency
\be
 \omega = \omega_R + i \omega_I
\ee
where $\omega_R$ and $\omega_I<0$ are determined by properties of {\it unstably} trapped null geodesics \cite{Yang:2012he}. For example, $\omega_R/m_\phi \approx -p_t/p_\phi$ where $p_t$, $p_\phi$ are the conserved momenta of a trapped null geodesic, while $\omega_I$ is determined by the rate at which nearby null geodesics move away from this trapped geodesic. $\omega_R$ is ${\cal O}(\ell)$ while $\omega_I$ is $O(1)$. 

We will do something similar for the wave equation in the spacetime (\ref{eq:metric}). It has been shown that the wave equation separates in this geometry \cite{Giusto:2004ip} so we will look for solutions of the form 
\begin{equation}
\Phi(t,z,r,\theta,\phi,\psi)=e^{-i\omega t+i\lambda z+i\mpsi \psi+i\mphi \phi}\Phi_r(r)\Phi_{\theta}(\theta). \label{eq:Psi}
\end{equation} 
where the angular harmonics $\Phi_\theta$ are labelled by an integer $\ell$. 

By analogy with the Kerr case just discussed, for large $\ell$ we expect there to exist quasinormal modes which are closely related to the trapped null geodesics. There are several important differences to the Kerr case. First, in the geometry (\ref{eq:metric}), the trapping is {\it stable} so we expect $\omega_I$ to be much smaller than in the Kerr case. Second, on ${\cal S}$, the trapped null geodesics have zero energy and KK momentum. Hence we expect to find quasinormal modes with $\lambda=0$ such that $\omega_R/\ell \approx 0$, i.e., $\omega_R$ does not scale with $\ell$.

We can also consider a null geodesic with tangent $V$ that does not lie on ${\cal S}$. Such geodesics have $-p_t = p_z$ so we would expect there to exist corresponding quasinormal modes with $\lambda \ne 0$ and $\omega \approx \lambda$. We will look for these modes by taking $\lambda = {\cal O}(\ell)$ and $\omega - \lambda = {\cal O}(1)$. 

We will determine quasinormal modes in two ways. For large $\ell$ we will use a matched asymptotic expansion inspired by a similar calculation in \cite{Chakrabarty:2015foa}. For general $\ell$ we will determine quasinormal modes numerically. For both methods we will need to use the ODEs resulting from separation of variables, which are  \cite{Giusto:2004ip} 
\begin{subequations}
 \begin{equation}
\frac{1}{\sin 2\theta}\frac{\mathrm{d}}{\mathrm{d}\theta}\left(\sin 2\theta\frac{\mathrm{d}\Phi_{\theta}(\theta)}{\mathrm{d}\theta}\right)+\left[A-\frac{\mpsi^2}{\cos^2\theta}-\frac{\mphi^2}{\sin^2\theta}+(\wb^2-\lb^2)\frac{a^2\eta}{R_z^2}(\cos^2\theta+n \cos 2\theta)\right]\Phi_{\theta}(\theta)=0 \label{eq:theq}
\end{equation} 
\begin{equation}
\frac{1}{r}\frac{\mathrm{d}}{\mathrm{d}r}\left[r(r^2+\alpha^2)\frac{\mathrm{d}\Phi_r(r)}{\mathrm{d}r}\right]+\left(\tilde{\kappa}^2r^2+1-\tilde{\nu}^2+\frac{\xi^2\,s^2}{r^2+\alpha^2}-\frac{\zeta^2s^2}{r^2}\right)\Phi_r(r)=0\,, \label{eq:radial}
\end{equation}
\label{eq:full}
\end{subequations}
where $A$ is a constant arising from the separation of variables and
\begin{align}
&\wb=\omega R_z,\qquad \lb=\lambda R_z,\qquad s=\frac{\sqrt{Q_1\,Q_2}}{R_z^2},\qquad  \alpha = s\,\sqrt{\eta},\qquad \tilde{\kappa} = \sqrt{\tilde{\omega}^2-\tilde{\lambda}^2}
\\
&\tilde{\nu} = \sqrt{1+A-\tilde{\kappa}^2\frac{Q_1+Q_2}{R_z^2}-(\tilde{\omega}-\tilde{\lambda})^2\frac{Q_p}{R_z^2}},
\\
&\xi = \sqrt{\eta}\left[\frac{\tilde{\omega}}{\eta}-\tilde{\lambda}\frac{Q_p(Q_1+Q_2)}{Q_1\,Q_2}+n\,m_\psi-m_\phi\,(n+1)\right],
\\
&\zeta = \sqrt{\eta}\left[\tilde{\lambda}+m_\psi\,(n+1)-n\,m_\phi\right]\,.
\label{eq:xizeta} \end{align} 

\subsection{Matched asymptotic expansion\label{sec:matching}}

We will look first for quasinormal modes corresponding to the null geodesics with tangent $V$ that are on, or near to, ${\cal S}$. On ${\cal S}$ these have $p_t=p_z=0$ and non-zero $p_\phi$, $p_\psi$ in general. Therefore we look for quasinormal modes with 
$|\mpsi|,\,|\mphi|\gg 1$ while keeping $\{\wb,\,\lb\}=O(1)$ in \eqref{eq:Psi}. Our aim is to solve the coupled system of equations (\ref{eq:full}) for the eigenvalue pair $\{A,\wb\}$. It turns out that if either $|m_\phi|$ or $|m_\psi|$ are large, the two eigenvalues essentially decouple. That is to say, one can first determine $A$ and a \emph{posteriori} determine $\wb$.

To see how this works in more detail, we start by looking at the angular equation (\ref{eq:theq}). In the $|\mpsi|,\,|\mphi|\rightarrow \infty$ limit, while keeping $\{\wb,\lb\}$ fixed, we can introduce the effect of $\wb$ and $\lb$ perturbatively. At leading order, we can ignore the term proportional to $\wb^2-\lb^2$ in \eqref{eq:theq}, so that it becomes the equation for spherical harmonics on $S^3$ with known eigenvalues $A=\ell (\ell +2)\equiv\mu_{\ell}^2$ where
\begin{equation}
\ell\geq |\mpsi|+|\mphi|,\;\;\;\ell \in \mathbb{Z}. \label{eq:l}
\end{equation}

From \eqref{eq:l}, $|\mpsi|,\,|\mphi|\rightarrow \infty$ is equivalent to taking $\ell\rightarrow\infty$ and $|\mpsi|,\,|\mphi|=O(\ell) $; we will work in this limit for simplicity in keeping track of the orders of various terms. The next order term in the large $\ell$ expansion will only affect the $\ell$ independent piece of $A$, that is to say, at large\footnote{This correction can be easily computed, but will not be needed in what follows. For the interested reader, when $\ell=|\mphi|+|\mpsi|$
$$
A\approx\mu_{\ell}^2+\left(n\frac{\mphi}{\ell}-(n+1)\frac{\mpsi}{\ell}\right)(\wb^2-\lb^2)\frac{a^2\eta}{R_z^2}+O(\ell^{-1})\,.
$$
} $\ell$
$$
A\approx\mu_{\ell}^2+O(1)\,.
$$
It turns out that we only need to know $A$ up to this order in $\ell$ to know the leading behaviour of the imaginary part of the quasinormal modes in this sector of perturbations.

We now turn our attention to the radial equation. Unlike the angular equation, we cannot use standard perturbation theory to determine $\tilde{\omega}$. Instead, we have to resort to a matched asymptotic expansion.

The radial equation (\ref{eq:radial}) can be written as
\begin{equation}
-y(y^2+s^2\eta)\frac{\mathrm{d}}{\mathrm{d} y}\left[{y(y^2+s^2\eta)\frac{\mathrm{d} \Phi_r}{\mathrm{d} y}}\right]+V(y) \Phi_r(y)=0 \label{eq:yeq}
\end{equation} 
where we introduce the dimensionless variable $y =r/R_z$ and define
\begin{equation}
V(y)=-\tilde{\kappa}^2y^6+a y^4-b y^2+c
\end{equation}
where
 $a=\ell^2a_0+\ell a_1+O(1)$, $b=\ell^2b_0+\ell b_1+O(1)$ and $c=\ell^2c_0+\ell c_1+O(1)$,
\begin{equation}
\begin{split}
a_0& =1,\,\;\;a_1=2\\b_0& =-s^2\eta+\frac{\mphi^2}{\ell^2}(2n+1)(1-j^2)s^2\eta,
\\ b_1&=-2s^2\eta+2\eta s^2\frac{\mphi}{l}\Big(\frac{\wb}{\eta}-\lb \frac{Q_p(Q_1+Q_2)}{\Qo\Qf}\Big)\left(nj-(n+1)\right)-2\lb s^2\eta \frac{\mphi}{l}\left(j(n+1)-n\right)\\c_0& =s^4\eta^2\frac{\mphi^2}{\ell^2}(j(n+1)-n)^2\;\text{ and }\;c_1=2s^4\eta^2\lb\frac{\mphi}{\ell }\left(j(n+1)-n\right). \label{eq:coeff}
\end{split}
\end{equation} 
For later use, we also define \begin{equation}
j\equiv\frac{\mpsi}{\mphi},\;\;m\equiv\frac{\mphi}{\ell}\,\,\Rightarrow |m|\leq \frac{1}{1+|j|}.
\end{equation}
The wave equation is invariant under complex conjugation and so we have an overall choice of sign in the exponent in \eqref{eq:Psi}. Geodesics with tangent vector $V$ on ${\cal S}$ have $\pphi<0$ so we will fix the sign by assuming $m<0$.

To calculate the frequencies of quasinormal modes we find solutions of \eqref{eq:yeq} obeying the necessary boundary conditions in the limit $\ell\rightarrow\infty$. We use an asymptotic matching procedure with $\ell\rightarrow\infty$ a large parameter, similar to that used in ref. \cite{Chakrabarty:2015foa} for the decoupling limit of non-supersymmetric 3-charge microstate geometries.

Note that $\{a,\,b,\,c\}=O(\ell^2)$ but $\tilde{\kappa}=O(1)$ so that we can split the $y-$axis into 3 regions, approximate the potential $V(y)$ and then solve the remaining equation exactly in each region. The regions and approximations of the potential are as follows:
\begin{enumerate}[1)]
	\item $\,y\ll \sqrt\ell $:  $V(y) \approx a y^4-b y^2+c$
	\item $1\ll y\ll\ell$: $V(y)\approx a y^4$
	\item $\,y\gg \sqrt{\ell}$:  $V(y)\approx -\tilde{\kappa}^2y^6+a y^4$.
\end{enumerate} Since region 2 overlaps with both regions 1 and 3 we can find solutions in each of the regions then match them where they overlap. We will label the solution of $\Phi_r$ in each of the regions by $\Phi_i$, where $i$ indexes the region in question.

\subsubsection{Region 1: $\,y\ll \sqrt\ell $}
We approximate the equation by
\begin{equation}
y(y^2+s^2\eta)\frac{\mathrm{d}}{\mathrm{d}y}\left[{y(y^2+s^2\eta)\frac{d \Phi_1}{d y}}\right]-(ay^4-by^2+c)\Phi_1(y)=0.
\label{eq:yeq1}
\end{equation}
To make the expressions more compact, we define
\begin{equation}
\alpha\equiv s\sqrt{\eta},\;\;\beta\equiv \sqrt{a+\frac{c}{\alpha^4}+\frac{b}{\alpha^2}},\;\; \nu\equiv \sqrt{1+a}=\ell +1+O(\ell^{-1}).
\end{equation}
Eq.~(\ref{eq:yeq1}) can be brought to a more familiar form by a suitable change of variable. We define
$$
\Phi_{1}(y)=y^{\frac{\sqrt{c}}{\alpha^2}}(y^2+\alpha^2)^{\frac{\beta}{2}}Q\left(-\frac{y^2}{\alpha^2}\right)\,,
$$
where we implicitly have changed to a new coordinate $\tilde{z}=-y^2/\alpha^2$. The resulting equation for $Q(\tilde{z})$ is that of a Gaussian hypergeometric function of the second kind, $_2 F_1(\tilde{a},\tilde{b},\tilde{c},\tilde{z})$ with
$$
\tilde{a}=\frac{1}{2}\left(1-\nu+\beta+\frac{\sqrt{c}}{\alpha^2}\right)\,,\quad \tilde{b} = \frac{1}{2}\left(1+\nu+\beta+\frac{\sqrt{c}}{\alpha^2}\right)\quad\text{and}\quad \tilde{c}= 1+\frac{\sqrt{c}}{\alpha^2}\,.
$$
Our boundary conditions demand that we choose the regular Gaussian hypergeometric function at $\tilde{z}=y=0$. Our final solution, in this region of the potential, can simply be written as
\begin{equation}
\Phi_{1}(y)=A_1 y^{\frac{\sqrt{c}}{\alpha^2}}(y^2+\alpha^2)^{\frac{\beta}{2}}\,_2F_1\left(\tilde{a},\tilde{b},\tilde{c},-\frac{y^2}{\alpha^2}\right).
\label{eq:H1}
\end{equation}
where $A_1$ is a constant.

To match to region 2 take the limit $y\rightarrow \infty$ ($\ell\rightarrow\infty$ and the overlap region is $1\ll y \ll \sqrt{\ell}$, so we can have for example $y\approx O(\ell^{\frac{1}{4}})$):
\begin{equation}
\Phi_1(y)\approx A_1\Gamma(1+\frac{\sqrt{c}}{\alpha^2})\alpha^{\frac{\sqrt{c}}{\alpha^2}+\frac{1}{2}+\frac{\beta}{2}}\Bigg[\alpha^{\frac{\nu}{2}}y^{-\nu -1} \frac{\Gamma(-\nu)}{\Gamma(\tilde{c}-\tilde{b})\Gamma(\tilde{a})} +\alpha^{-\frac{\nu}{2}}y^{\nu -1} \frac{\Gamma(\nu)}{\Gamma(\tilde{c}-\tilde{a})\Gamma(\tilde{b})}\Bigg].
\label{eq:1H2}
\end{equation}

\subsubsection{Region 2: $1\ll y\ll\ell$}
In this region the equation is approximated by \begin{equation}
y^3\frac{d}{dy}\left(y^3\frac{d\Phi_2}{dy}\right)-a y^4 \Phi_2(y)=0 \label{eq:yeq2}
\end{equation} since $s^2\eta\ll y^2$. This has solution \begin{equation}
\Phi_{2}(y)=B_1 y^{-\nu-1}+B_2y^{\nu -1} \label{eq:H2}
\end{equation} where $B_1,\,B_2$ are constants. 

Matching \eqref{eq:1H2} to \eqref{eq:H2} in the overlapping region gives the condition: \begin{equation}
\frac{B_1}{B_2}=\alpha^{\nu}\frac{\Gamma(-\nu)}{\Gamma(\nu)}\frac{\Gamma(\tilde{c}-\tilde{a})\Gamma(\tilde{b})}{\Gamma(\tilde{c}-\tilde{b})\Gamma(\tilde{a})}.
\label{eq:b1b2}
\end{equation}

\subsubsection{Region 3: $y\gg \sqrt{\ell}$}
In this region at highest order in $\ell $, \begin{equation}
y^3\frac{d}{dy}\left(y^3 \frac{d\Phi_3}{dy}\right)-\left(-\tilde{\kappa}^2y^6+ay^4\right)\Phi_3=0 \label{eq:yeq3}
\end{equation}
with solution \begin{equation}
\Phi_3(y)=\frac{1}{y}\big(C_1J_\nu(\tilde{\kappa} y)+C_2Y_\nu(\tilde{\kappa} y)\big) \label{eq:H3}
\end{equation}
where $C_1,\,C_2$ are constants and $J_\nu(x),\,Y_\nu(x)$ are Bessel functions of the first and second kind respectively.

In the asymptotic region as $y\rightarrow \infty$,
\begin{equation}
\Phi_3(y)= \frac{1}{y^{\frac{3}{2}}}\frac{1}{\sqrt{\tilde{\kappa}\pi}}\left[ e^{i\tilde{\kappa} y} e^{-i\frac{\nu \pi}{2}}\left(\frac{1}{2}-\frac{i}{2}\right)(C_1-iC_2)+ e^{-i\tilde{\kappa} y} e^{i\frac{\nu \pi}{2}}\left(\frac{1}{2}+\frac{i}{2}\right)(C_1+iC_2)\right]+O(y^{-\frac{5}{2}}).
\label{eq:h3inf}
\end{equation}
Imposing the boundary condition that there are only outgoing waves at infinity gives
\begin{equation}
C_1+iC_2=0.
\label{eq:infcond}
\end{equation}

To match to Region 2 in the overlap region $\sqrt{\ell}\ll y\ll\ell$ we take $\tilde{\kappa} y\ll\ell$ while $\nu\rightarrow\infty$. Using the formulae for the asymptotic form of the Bessel functions at large orders \cite{AS} gives:
\begin{equation} 
\Phi_3= \left[C_1y^{\nu-1}\left(\frac{\tilde{\kappa}}{2}\right)^{\nu}\frac{1}{\sqrt{2\pi\nu}}\frac{e^{\nu}}{\nu^{\nu}}-C_2y^{-\nu-1}\left(\frac{\tilde{\kappa}}{2}\right)^{-\nu}\sqrt{\frac{2}{\pi \nu}}\frac{e^{-\nu}}{\nu^{-\nu}}\right]\left[1+O(\ell^{-1})\right]
\label{eq:3H2}
\end{equation}
and so we find
\begin{equation}
\frac{C_1}{C_2}=-2\left(\frac{\tilde{\kappa}}{2}\right)^{-2\nu}e^{2\nu}\nu^{-2\nu}\frac{B_2}{B_1}.
\label{eq:c1c2}
\end{equation}

\subsubsection{Real part of the frequency}\label{sec:realw}
The conditions \eqref{eq:b1b2}, \eqref{eq:infcond} and \eqref{eq:c1c2} all together imply that the quasinormal mode frequencies $\wb$ are solutions of the equation
\begin{equation}
\alpha^{-\nu}\frac{\Gamma(\tilde{c}-\tilde{b})\Gamma(\tilde{a})}{\Gamma(\tilde{c}-\tilde{a})\Gamma(\tilde{b})}=2i\left(\frac{\tilde{\kappa}}{2}\right)^{2\nu}\frac{ \Gamma(-\nu)}{\Gamma(\nu)}e^{2\nu}\nu^{-2\nu}. \label{eq:star}
\end{equation}
We have that $\nu=\sqrt{1+a}=O(\ell) \gg 1$ so the RHS is extremely small; the only way to solve \eqref{eq:star} is to have a pole in one of the $\Gamma-$functions in the denominator of the LHS \emph{i.e.}
\begin{equation}
(\tilde{c}-\tilde{a}=-N\lor \tilde{b}=-N) \Rightarrow \frac{1}{2}\left(1+\nu\pm\beta+\frac{\sqrt{c}}{\alpha^2}\right)=-N.
\label{eq:pole}
\end{equation}

The leading order dependence on $\wb$ in \eqref{eq:pole} comes from
\begin{equation}
\beta=\ell|m(jn-(n+1))|+\frac{|m(nj-(n+1))|}{m(nj-(n+1))}\left(\frac{\wb}{\eta}-\lb\frac{Q_p(\Qo+\Qf)}{\Qo\Qf}\right)+O(\ell^{-1}).
\end{equation}

From the condition that $|\wb|, |\lb| \ll\ell$, all the terms that are proportional to $\ell $ in \eqref{eq:pole} must cancel:
\begin{equation}
1+|m(j(n+1)-n)|\pm |m(jn-(n+1))|=0. \label{eq:lcancel}
\end{equation}

Clearly, this condition does not hold for general values of $m$ and $j$, and so we will use \eqref{eq:lcancel} to find possible values for $m$ in terms of $j$ for which there are quasinormal modes with $|\wb|, |\lb| \ll\ell$. By examining \eqref{eq:lcancel} we see that it can only be solved if we choose the minus sign (otherwise all terms on the left hand side are positive). The  equation remains non-trivial. We will use geometric optics to help us find a solution. 

In geometric optics, $j=p_\psi/p_\phi$ to leading order in $\ell$. In section \ref{sec:geo} we found that the zero energy geodesics with tangent vector $V$ have:
\be
0\leq \frac{\ppsi}{\pphi}\leq\frac{n}{n+1}\quad \text{for}\quad n>0, \qquad \frac{\ppsi}{\pphi}\geq\frac{n}{n+1}\quad \text{for}\quad n<0 
\ee
This suggests that we look for a solution of  \eqref{eq:lcancel} with 
\be
\label{jrange}
0\leq  j\leq\frac{n}{n+1}\quad \text{for}\quad n>0, \qquad j \geq\frac{n}{n+1}\quad \text{for}\quad n<0 
\ee
In both cases we have $j \ge 0$ and $(n+1)j-n\leq 0$, and these imply $nj -(n+1) < 0$. Using these, along with $m<0$, equation \eqref{eq:lcancel} reduces to
\begin{equation}
m=-\frac{1}{1+j} \label{eq:m}
\end{equation} 
which is equivalent to
\be
\label{ellsol}
 \ell = -m_\phi-m_\psi
\ee 
So in summary, we have found values of $\ell$, $m_\phi$, $m_\psi$ that are consistent with our assumptions by taking $m_\phi,m_\psi<0$ and $j=m_\psi/m_\phi$ in the range \eqref{jrange}, with $\ell$ given by \eqref{ellsol}. Substituting these values into \eqref{eq:pole}, the real part of $\wb$ at leading order is \begin{equation}
\wb_R=2\eta (N+1)+\lb. \label{eq:wr}
\end{equation}	
The expression \eqref{eq:wr} for $\wb_R$ is remarkably simple. As a check on this formula we can take the decoupling limit $Q_p\ll \sqrt{\Qo\Qf}\ll R_z^2$, which gives $\eta\rightarrow 1$, in \eqref{eq:wr}. In this limit the geometry reduces to AdS$_3\times S^3$ and our expression for $\wb_R$ reduces to the formula for certain normal modes in AdS$_3\times S^3$, see e.g. Eq.~(6.12) of \cite{Giusto:2012yz}.\footnote{A similarly simple expression was found for the real part of the frequencies of {\it unstable} modes in the non-supersymmetric 3-charge geometries in the decoupling limit in \cite{Chakrabarty:2015foa} although in that case the real part of the frequency scales as $\ell$ in general.}
	
\subsubsection{Imaginary part of the frequency\label{sec:imw}}	
To find the imaginary part of the frequency we look at the next order terms in \eqref{eq:star} by substituting $\wb=\wb_R+\delta\wb$. Then $\beta=\beta(\wb_R)+\delta \beta$ where $\delta\beta=\frac{\delta\wb}{\eta}$ and we substitute
\begin{equation}
\Gamma(-N-\frac{\delta\beta}{2})=\frac{(-1)^{N+1}}{N!}\frac{2}{\delta \beta}(1+O(\delta\beta))
\end{equation}
in the left hand side of \eqref{eq:star}, which is the only term that depends on $\delta \beta$ at highest order. We also use the well known identities
\begin{equation*}
\Gamma(-\nu)=-\frac{\pi}{\nu \sin \pi\nu}\frac{1}{\Gamma(\nu)}\,,\quad\text{and}\quad\Gamma(-N-\nu)=\frac{(-1)^{N+1}\pi}{(N+\nu)\sin\pi\nu}\frac{1}{\Gamma(N+\nu)}.
\end{equation*}
Substituting these into \eqref{eq:star} and rearranging: \begin{equation}
\delta \beta =-i\left(\frac{\tilde{\kappa}}{2}\right)^{2\nu}\alpha^{\nu}\frac{4 (N+\nu) \Gamma(N+1+\nu+\frac{\sqrt{c}}{\alpha^2})}{N!\nu\Gamma(N+1+\frac{\sqrt{c}}{\alpha^2})}\frac{\Gamma(N+\nu)}{\Gamma(\nu)^2}e^{-2\nu\log \nu+2\nu}. \label{eq:db}
\end{equation}
The size of the corrections to the real part of the frequency $\wb_R$ from Eq.~\eqref{eq:pole} are of order $O(\ell^{-1})$ and are thus much larger than the corrections to $\wb$ here. However, the corrections to $\wb$ in \eqref{eq:pole} will all be real (all the coefficients are real apart from dependence on $\wb$) and so the imaginary part of the frequency does not have any terms that are proportional to inverse powers of $\ell$. We therefore use \eqref{eq:db} to find the imaginary part of $\wb$ at leading order and we in fact have $\delta \wb=\delta\wb_R+i\wb_I$. Substituting this in to \eqref{eq:db}, we find
\begin{equation}
 \wb_I=-\eta\left(\frac{\tilde{\kappa}}{2}\right)^{2\nu}\alpha^{\nu}\frac{4 (N+\nu) \Gamma(N+1+\nu+\frac{\sqrt{c}}{\alpha^2})}{N!\nu\Gamma(N+1+\frac{\sqrt{c}}{\alpha^2})}\frac{\Gamma(N+\nu)}{\Gamma(\nu)^2}e^{-2\nu\log \nu+2\nu}. \label{eq:wi}
\end{equation}
Define $\mu=-\frac{j(n+1)-n}{1+j}>0$, then use $\ell\gg 1$ in \eqref{eq:wi} gives \begin{equation}
\wb_I=-D \eta \alpha \tilde{\kappa}_0^2 e^{-2\ell\log\ell+\ell\left[2-\mu\log\mu+(1+\mu)\log(\mu+1)+2\log\frac{\tilde{\kappa}_0\sqrt{\alpha}}{2}\right]+(N-\frac{3}{2})\log \ell+O(1)} \label{eq:wi2}
\end{equation}
where $\tilde{\kappa}_0=\sqrt{\wb_{R,0}^2-\lb^2}$, $\wb_{R,0}$ is the real part of $\wb$ calculated to leading order only (i.e. $\wb_R$ in \eqref{eq:wr}). $D$ is a positive constant that is independent of $\ell$ at leading order but depends on the higher order corrections to the real part of $\wb$ from the term $\tilde{\kappa}^{2(\ell+1)}$ in \eqref{eq:wi}. 

Equation \eqref{eq:wi2} is one of our main results. We see that $\wb_I<0$ so the waves decay as expected. However, the rate of decay is very slow, since at leading order the term that controls it is $e^{-2\ell  \log \ell}$ which is very small for large $\ell$. 

As discussed above, in the decoupling limit we expect our quasinormal modes to reduce to normal modes in AdS$_3 \times$ S$^3$ so $\wb_I$ should vanish in this limit. This is indeed the case because $\alpha \rightarrow 0$ in the decoupling limit. 

The calculation above assumes $n \ne 0$, i.e., $Q_p \ne 0$ so it does not apply to 2-charge microstate geometries. When $n=0$, ${\cal S}$ becomes the 2-dimensional submanifold $r=0$, $\theta=\pi/2$. In Appendix \ref{app:2charge} we show that it is straightforward to modify the above calculation to cover this case too. The result is the same, i.e, $\wb_I$ is ${\cal O}(e^{-2\ell \log \ell})$ at large $\ell$. Hence the dimension of ${\cal S}$ does not seem to affect the slow decay, which is to be expected since the slowly decaying quasinormal modes are associated to individual null geodesics on ${\cal S}$ rather than to global properties of ${\cal S}$.


\subsection{Kaluza-Klein momentum scaling with $\ell$}

In section \ref{sec:geo} we saw that at every point in the six-dimensional spacetime there is a stably trapped geodesic with tangent $V$. We have found quasinormal modes that correspond to the zero energy null geodesics that are trapped near $\mathcal{S}$ but we also expect to be able to find slowly decaying modes that are localised near geodesics that are trapped elsewhere in the spacetime. These geodesics have tangent $V$ and conserved quantities $p_z=-p_t$. Under the geometric optics approximation we expect that the corresponding solutions of the wave equation will have $\wb \approx \lb$. We will now consider $\lb = {\cal O}(\ell)$ but keep the difference $|\wb-\lb|=O(1)$ in the limit $|\mpsi|,\,|\mphi|\rightarrow\infty$. In this case, $\tilde{\kappa}^2=(\wb-\lb)(\wb+\lb)=O(|\mpsi|,\,|\mphi|)$. 

Since $\tilde{\kappa}^2\ll \mphi^2,\;\mpsi^2$, we can ignore the $\tilde{\kappa}^2$ in the angular equation \eqref{eq:theq} at leading order in $m_\phi$, $m_\psi$. This means that we have
$$
A\approx \ell^2+A_1\ell+O(1)
$$
with $\ell$ defined previously in \eqref{eq:l}. If we set $\ell=|\mphi|+|\mpsi|$, \emph{i.e.} $m=-1/(1+j)$, we can find $A_1$ using standard perturbation theory. It turns out that 
\be
A_1=2-2\frac{\lb\alpha^2}{\ell}(\wb-\lb)\left(\frac{n-(n+1)j}{1+j}\right)
\ee
We will find later that we must have $m=-1/(1+j)$ to have modes $|\wb-\lb|=O(1)$ so this assumption is consistent.

The expressions for $a,\;b,\;c$ in \eqref{eq:yeq} at the various orders change: we now have
\begin{equation} \begin{split}
& a=\tilde{\nu}^2-1-\tilde{\kappa}^2\alpha^2 \\ 
& b=\alpha^2(1-\tilde{\nu}^2)+s^2(\xi^2-\zeta^2) \\
& c=\alpha^2s^2\zeta^2
\end{split} \end{equation} where $\tilde{\nu},\;\xi,\;\zeta$ are defined in \eqref{eq:xizeta}.

\subsubsection{Asymptotic matching}\label{sec:matchingell}
The asymptotic matching procedure in \ref{sec:matching} only needs to be slightly modified to find solutions with frequencies with $\tilde{\kappa}^2=O(\ell)$. Regions 1, 2 and 3 must be changed so that the potentials can be approximated in the same way as before in each region.

We define the new regions as:
\begin{enumerate}[1')]
	\item $\,y\ll \ell^{\frac{1}{4}} $: $V(y)\approx a y^4-by^2+c$ 
	\item $1\ll y\ll\sqrt{\ell}$: $V(y)\approx a y^4$;
	\item $\,y\gg \ell^{\frac{1}{4}}$:  $V(y)\approx -\tilde{\kappa}^2y^6+a y^4$.
\end{enumerate}  Note that the regions still overlap so we can match the solutions in different regions.

Exactly the same matching procedure as in section \ref{sec:matching} then follows through to give that the real part of the frequency is defined by the condition \begin{equation}
\frac{1}{2}(1+\nu\pm\beta+\frac{\sqrt{c}}{\alpha^2})=-N. \label{eq:pole2}
\end{equation}
We expect $\wb-\lb$ to be small so we must take the minus sign for the same reasons as in section \ref{sec:realw}. However, the leading order behaviour of $\beta$ and $\nu$ differs to the previous case; we find that now
\begin{equation}
\begin{split}
&\beta=\lb+m\ell[jn-(n+1)]+(\wb-\lb)\left\{\frac{1}{\eta}-\frac{\lb\alpha^2}{\lb+m\ell[jn-(n+1)]}\right\}+O(\ell^{-1})\\
& \frac{\sqrt{c}}{\alpha^2}=\lb+m\ell[j(n+1)-n],\\
& \nu=\ell+\frac{A_1}{2}-(\wb-\lb)\frac{\lb}{\ell}\left(\frac{\Qo+\Qf}{R_z^2}+\alpha^2\right)+O(\ell^{-1})=\ell+\nu_1+O(\ell^{-1}).
\label{eq:bcn}
\end{split}
\end{equation}
We assume as before that $m<0,\;jn-(n+1)<0,\;j>0$ and $\lb\geq0$. Substituting this into equation \eqref{eq:pole2} and imposing the condition $|\wb-\lb|=O(1)$, we find that we must take $m=-1/(1+j)$ so that the higher order terms cancel. Then the real part of the frequency is given by \eqref{eq:pole2}:
\begin{equation}
\wb_R=\lb+\frac{2\eta}{P}(N+1) +O(\ell^{-1}) \label{eq:wr2}
\end{equation} where we use the definitions of $a,\;b,\;c$ and $\wb=\lb+O(1)$ to find
\begin{equation}
P=1+\frac{\lb\alpha^2\eta}{\ell}\left(1-\frac{\ell}{\lb+m\ell[jn-(n+1)]}\right)+\frac{\lb}{\ell}\eta\frac{\Qo+\Qf}{R_z^2}+ \frac{\lb\alpha^2\eta}{\ell}\left(\frac{n-j(n+1)}{1+j}\right)>0.
\end{equation}
If we take $\lb\ll \ell$ in \eqref{eq:wr2} we recover the real part of the frequency for $\lb=O(1)$ as given in eq. \eqref{eq:wr}.

The calculation for the imaginary part is also very similar to that of section \ref{sec:imw}; we simply have to replace $\delta \beta$ with $P \delta\wb$ in \eqref{eq:wi}. Then let
\begin{equation*}
\mu'=\frac{\tilde{\lambda}}{\ell}-\frac{j(n+1)-n}{1+j}>0.
\end{equation*}
In the limit $\ell\rightarrow\infty$, the imaginary part of the frequency at leading order is 
\begin{equation}
\wb_I=-D'e^{-\ell\log\ell+\ell\left[2-\mu'\log\mu'+(1+\mu')\log(\mu'+1)+2\log\frac{\tilde{\kappa}_0\sqrt{\alpha}}{2\sqrt{\ell}}\right]+(N+\frac{1}{2}-\nu_1)\log \ell} +O(l^{-1}) \label{eq:wi2l}
\end{equation}
for some positive constant $D'$ that is independent of $\ell$ and where $\tilde{\kappa}_0^2=2\lb(\wb_R-\lb)$ with $\wb_R$ evaluated using \eqref{eq:wr2} and $\nu_1$ is given in \eqref{eq:bcn} with the terms $\wb-\lb$ also evaluated at leading order using \eqref{eq:wr2}. $D'$ is proportional to $\alpha^{\nu_1}$; in the decoupling limit $\nu_1\rightarrow 1$ and $ \alpha\rightarrow 0$ so we see that the imaginary part vanishes in this limit, as expected. The real part reduces to the expression for certain normal modes in AdS$_3\times S^3$, as given in \cite{Giusto:2012yz}.


We have constructed quasinormal modes with $ \wb_I \sim -e^{-\ell\log \ell}$ at leading order for $\ell \gg 1$. We expect that such a mode will be localised near a stably trapped geodesics with tangent $V$, whose location is determined by the matching the ratios $p_\psi/p_\phi$, $p_z/p_\phi$ to $m_\psi/m_\phi$ and $\lambda/m_\phi$. Note that there is no longer a factor of $2$ multiplying $-\ell\log\ell$ in the exponent so these modes decay faster than the modes localized near ${\cal S}$ that we found in the previous section. However, the decay is still very slow and therefore likely to be problematic for nonlinear stability. 

The above calculation assumes $n\ne 0$, i.e., $Q_p \ne 0$ but in Appendix \ref{app:2charge} we show that it is straightforward to modify the calculation to cover the 2-charge case. The result is $\wb_I = {\cal O}(e^{-\ell \log \ell})$ as for the 3-charge case. 

\subsection{Numerical determination of quasinormal modes}

\subsubsection{Method}

In the previous sections we have determined certain quasinormal modes in the limit of large quantum number $\ell$, we now aim to determine the behaviour of the corresponding modes at finite $\ell$ numerically. In doing so, we can also understand the regime of validity of the approximation scheme detailed in our previous sections. For the sake of presentation, we will restrict ourselves to the case with $\tilde{\lambda}=0$, \emph{i.e.} modes that do not depend on the Kaluza-Klein momentum.

Our separation ansatz reads
$$
\Phi(r,\theta) = X(\cos \theta)\,W\left(\frac{r R_z}{\sqrt{Q_1\,Q_2}}\right)\,,
$$
which yields the following pair of ordinary differential equations for $X(x)$ and $W(w)$ to be solved numerically:
\begin{subequations}
\begin{multline}
\frac{1}{x}\frac{\mathrm{d}}{\mathrm{d}x}\left[x\,(1-x^2)\frac{\mathrm{d}}{\mathrm{d}x} X(x)\right]+\\
\left\{A+\alpha _1 \alpha _2  \eta  \tilde{\omega }^2 \left[-n \left(1-x^2\right)+x^2 (1 +n)\right]-\frac{m_{\psi }^2}{x^2}-\frac{m_{\phi }^2}{1-x^2}\right\}X(x)=0
\label{eq:angular}
\end{multline}
\begin{multline}
\frac{1}{w}\frac{\mathrm{d}}{\mathrm{d}w}\left[w\,(w^2+\eta )\frac{\mathrm{d}}{\mathrm{d}w} W(w)\right]+\\
\Bigg\{\tilde{\omega }^2
   \left[\alpha _1+\alpha _2+\alpha _1 \alpha _2 n (1+n)+\alpha _1 \alpha _2 w^2\right]-A-\frac{\eta  \left[n m_{\phi }-(1+n) m_{\psi }\right]^2}{w^2}+\\
   \frac{\eta  \left[\left(\alpha _1+\alpha _2\right) n (1+n) \tilde{\omega }+\tilde{\omega }+n m_{\psi }-(1+n) m_{\phi }\right]^2}{w^2+\eta }\Bigg\}W(w)=0\,,
   \label{eq:radial2}
\end{multline}
\label{eq:eqs}
\end{subequations}
where we have changed variables to $x\equiv \cos \theta$ and $w \equiv r R_z/\sqrt{Q_1\,Q_2}$ and defined $Q_i = \alpha_i \, R_z^2$. Here, as in previous sections, $A$ is a separation constant to be determined in what follows.

Before detailing our numerical method, we need to investigate the boundary conditions at the edges of our integration domain. Our equations have five real singular points (three for the angular equation, and two for the radial equation). For the angular equation (\ref{eq:angular}) these are $x=0$, $x=1$ and $x=\infty$. For the radial equation these are $w=0$ and $w=\infty$.

Let us start with the angular equation. Since our integration domain is $x\in(0,1)$, we only need to understand what happens at these singular points. A Frobenius expansion at $x=0$, yields the following behaviour
$$
X\sim x^{\pm |m_\psi|}\left[1+\mathcal{O}(x)\right]\,,
$$
while at $x=1$ we find
$$
X\sim (1-x)^{\pm \frac{|m_\phi|}{2}}\left[1+\mathcal{O}(1-x)\right]\,.
$$
In order to have a regular solution, we must choose the $+$ signs at both integration edges. To solve the problem numerically, we change to a new variable that relates to $X$ in the following manner:
$$
X=x^{|m_{\psi}|}(1-x^2)^{\frac{|m_\phi|}{2}}\,\widetilde{X}\,,
$$
and impose Robin boundary conditions for $\widetilde{X}$ at $x=0$ and $x=1$. These can be found by solving the equations for $\widetilde{X}$ in a Taylor expansion around the two singular points.

Next we address the radial equation. The singular point at $w=0$ is a regular singular point, and its behaviour can be extracted via a Frobenius expansion (similar to the angular equation):
$$
W(w) \sim w^{\pm|n (m_{\phi }-m_{\psi})-m_{\psi }|}\left[1+\mathcal{O}(w)\right]\,,
$$
again regularity demands keeping the $+$ sign only. Finally, at $w=+\infty$, there is an essential singularity, which is to be expected since we want to impose outgoing boundary conditions there. The singular behaviour can be easily extracted, and takes the following form
$$
W(w)\sim \frac{e^{\pm i \sqrt{\alpha _1} \sqrt{\alpha _2} w \tilde{\omega }}}{w^{\frac{3}{2}}}\left[1+\mathcal{O}(w^{-1})\right]\,.
$$
Demanding outgoing boundary conditions yields demands choosing the $+$ sign. As we have done for the angular equation, we now change to a new variable that is more adequate for the numerical procedure. We chose the following:
$$
W(w)=\frac{e^{i \sqrt{\alpha _1} \sqrt{\alpha _2} w \tilde{\omega }}}{w^{\frac{3}{2}+|n (m_{\phi }-m_{\psi})-m_{\psi }|}}w^{|n (m_{\phi }-m_{\psi})-m_{\psi }|}\widetilde{W}(w)\,.
$$
Finally, since $w$ is a non-compact coordinate, we do a further change of coordinates of the form
$$
w=\frac{\tilde{w} \sqrt{2-\tilde{w}^2}}{1-\tilde{w}^2}\,,
$$
which maps $w=0$ to $\tilde{w}=0$ and $w=+\infty$ to $\tilde{w}=1$. Robin boundary conditions at $\tilde{w}=0$ and $\tilde{w}=1$ can now be found by solving the respective equation for $\widetilde{W}(\tilde{w})$.

Our original system of equations (\ref{eq:eqs}) has been mapped to two equations for $\widetilde{X}(x)$ and $\widetilde{W}(\tilde{w})$, with two coupled eigenvalues $(\tilde{\omega},A)$. In order to solve these, we use a Newton-Raphson routine which has been outlined in \cite{Cardoso:2013pza} for a similar problem. Regarding the implementation of the algorithm, the only nuance that is worth emphasising is that we had to work with arbitrary-precision arithmetic, since the magnitude of the imaginary part of our quasinormal modes can be as small as $10^{-170}$ (for an order of magnitude, this is more than the number of decimal places captured by octuple precision - $10^{-71}$).

\subsubsection{Results}

We have varied parameters in our search, \emph{i.e.} different values of $n$, $\alpha_i$, but the results look qualitatively similar. We divide the types of quasinormal modes we find into two types: $i)$ those for which $\omega_R$ does not scale with $\ell$ and $ii)$ those for which $\omega_R$ does scale with $\ell$. In this paper we will focus on type $i)$ modes, which is the sector that is responsible for the slow decay of generic perturbations. As we have seen in section \ref{sec:realw} (see discussion around Eq.~(\ref{eq:m})), the slow decay will only hold for modes satisfying $\ell=|m_\phi|+|m_\psi|$, which are the modes we are going to focus on.

For the sake of presentation, we will keep $\alpha_1=\alpha_2=1=n=1$. Changing $\alpha_1$ or $\alpha_2$ will just change the regime at which the matched asymptotic expansion analysis settles in. The larger $\alpha_1$ or $\alpha_2$, the larger the value of $\ell$ we need to reach in order to see matching with the matched asymptotic expansion analysis of the previous sections.

\begin{figure}[h]
\centering
\includegraphics[width=0.9\linewidth]{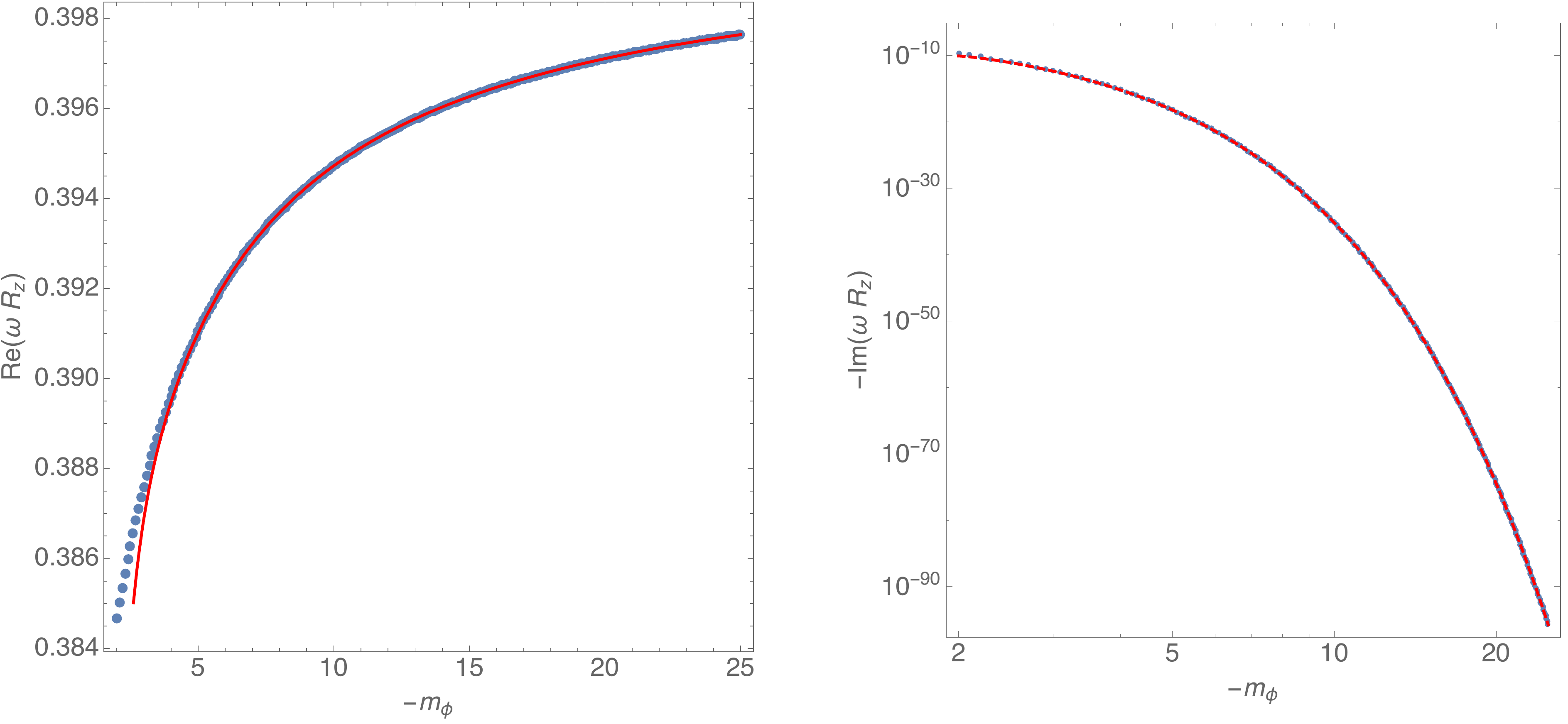}
\caption{\emph{Left panel}: real part of $\tilde{\omega}$ as a function of $m_\phi<0$. \emph{Right panel}: imaginary part of $\tilde{\omega}$ as a function of $m_\phi$. In both panels, the blue points are the numerical data, the solid red line is the analytic prediction for $\mathrm{Re}(\tilde{\omega})$ based on a quasimode construction (see Appendix \ref{app:quasi}), the dashed red line is the fit to (\ref{eq:wi2}), and both plots were generated with $\alpha_1=\alpha_2=n=-m_\psi=1$.}
\label{fig:n1}
\end{figure}

In Fig.~\ref{fig:n1} we show a linear plot (left panel) of the real part of $\tilde{\omega}$ as a function of $m_\phi<0$ for $m_\psi=-1$. We see that as $|m_\phi|$ increases, $\tilde{\omega}_R$ approaches the value predicted in Eq.~(\ref{eq:wr}). The approach to this value (solid red curve) can also be determined via the explicit construction of "quasimodes", which we detail in Appendix \ref{app:quasi}. On the right panel of the same figure, we show a log-log plot of the imaginary part of $\tilde{\omega}$ as a function of $|m_\phi|$: the blue dots are the numerical data, whereas the red dashed curve is a one parameter fit to (\ref{eq:wi2}), with $D$, the overall scale, being the fitting parameter.

The agreement of the fit with the numerical data is very reassuring. In fact, the agreement is much better than one might have expected: our analytical result \eqref{eq:wi2} works well down to small values of $\ell$ whereas this result was only expected to hold for $\ell \gg 1$. Note that the imaginary part of $\tilde{\omega}$ is very small even for small values of $\ell$. So there exist very slowly decaying quasinormal modes even at small $\ell$. The decay becomes even slower at high $\ell$, in agreement with our analytical result.

\begin{figure}[h]
\centering
\includegraphics[width=0.9\linewidth]{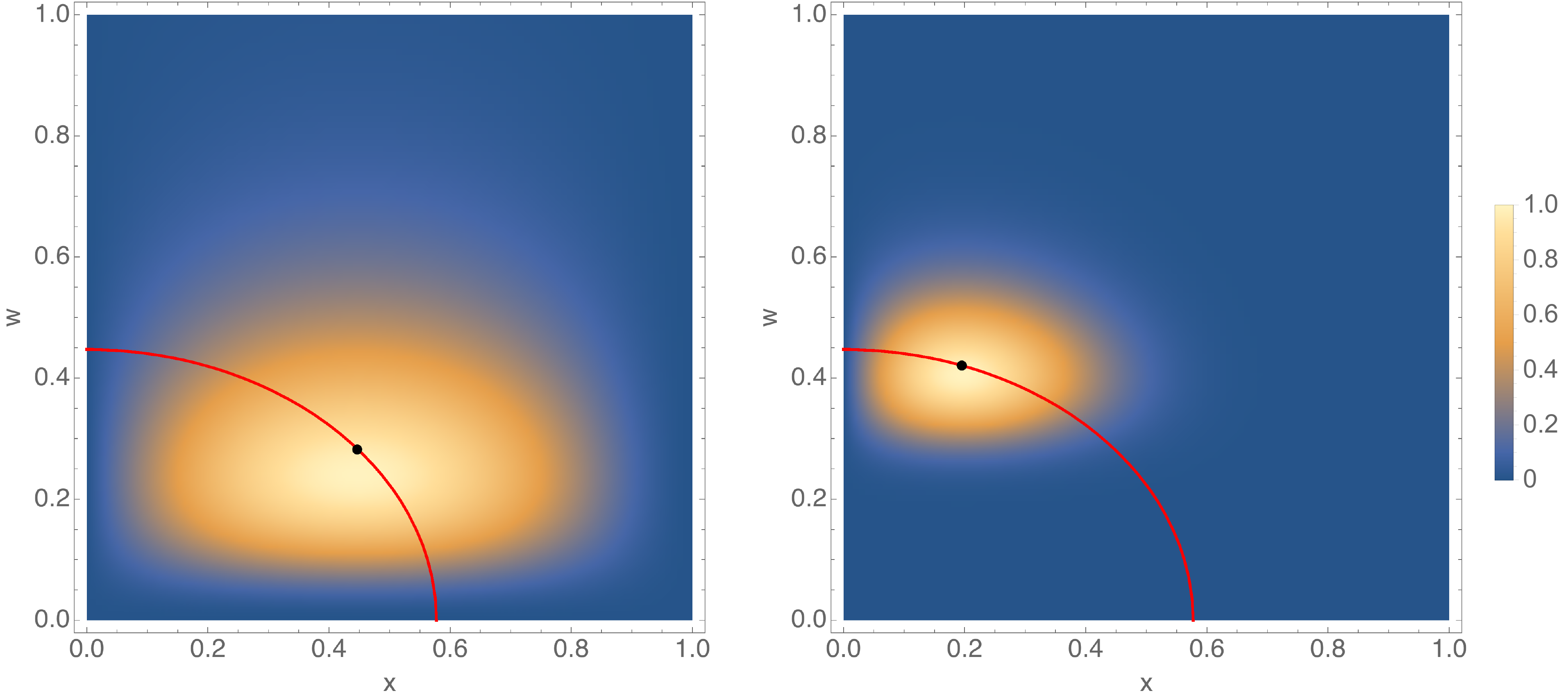}
\caption{Contour plot for $|\Phi|$ as a function of $w$ and $x$ on a surface extending to future null infinity. The red curve is the evanescent ergosurface ${\cal S}$. On the \emph{left panel} we have $m_\phi=-4$ and on the \emph{right panel} we have $m_\phi=-25$. Both panels were generated with $\alpha_1=\alpha_2=n=-m_\psi=1$ and the normalization is $\max |\Phi|=1$.}
\label{fig:n2}
\end{figure}

Quasinormal modes grow exponentially at spatial infinity but they are well behaved at future null infinity. We can consider the behaviour of quasinormal modes on a surface of constant retarded time $u=t-r$, which extends to future null infinity. In Fig.~\ref{fig:n2} we plot the absolute value of the quasinormal mode as a function of $w$ and $x$ on such a surface for the smallest and largest value of $m_\phi$ we studied. The idea is to see if the quasinormal mode is localized near the corresponding null geodesic on ${\cal S}$, i.e., the geodesic with $p_\psi/p_\phi=m_\psi/m_\phi$ (represented in Fig.~\ref{fig:n2} by a black dot). We see that as $m_\phi$ increases, the maximum moves towards $x=0$, as a consequence of the fact that $m_\phi$ is increasing, while $m_\psi$ is kept constant, so the ratio $m_\psi/m_\phi$ decreases. Furthermore, the quasinormal mode localises more sharply around the geodesic prediction, as expected from geometric optics because $\ell=|m_\phi|+|m_\psi|$ is increasing. 

\begin{figure}[h]
\centering
\includegraphics[width=0.9\linewidth]{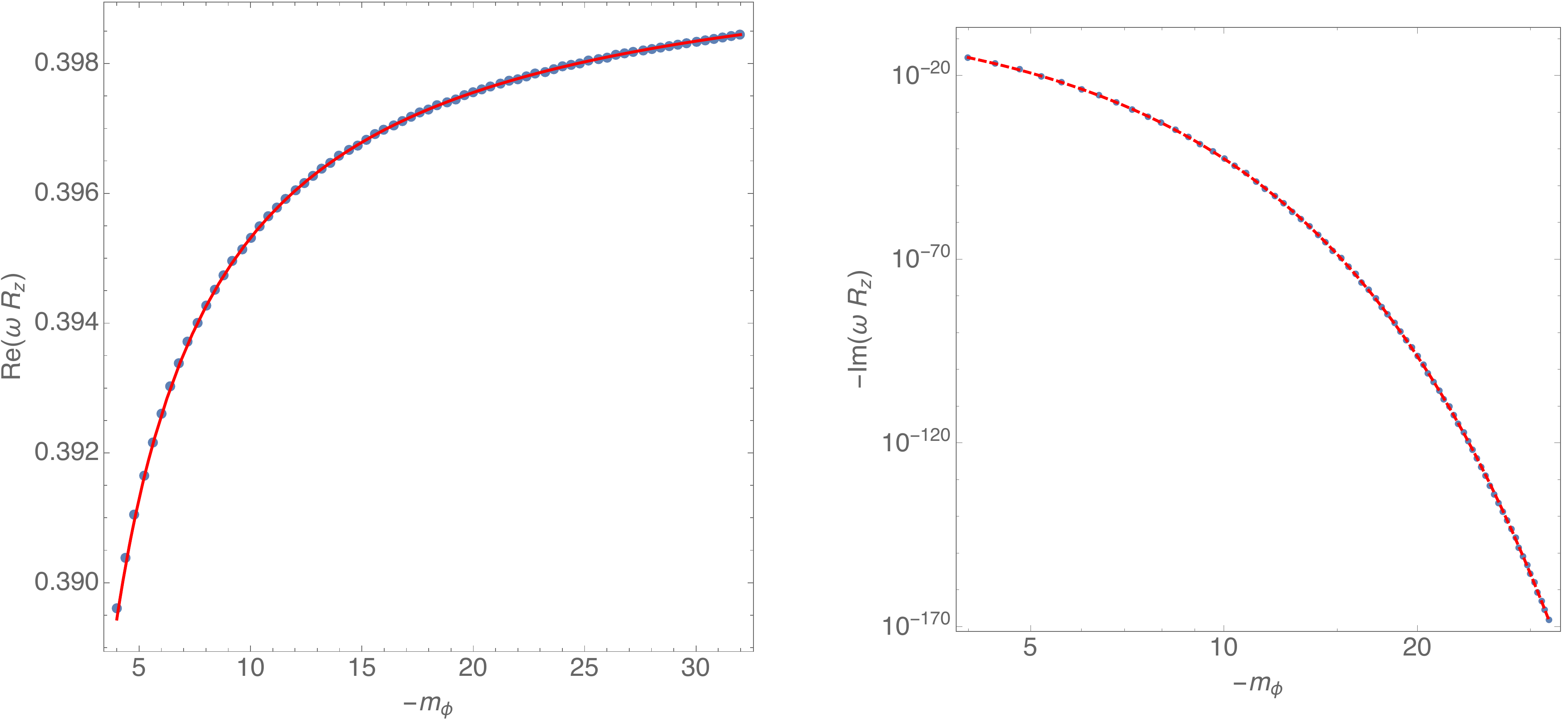}
\caption{\emph{Left panel}: real part of $\tilde{\omega}$ as a function of $m_\phi<0$. \emph{Right panel}: imaginary part of $\tilde{\omega}$ as a function of $m_\phi$. In both panels, the blue points are the numerical data, the dashed red line is the fit to (\ref{eq:wi2}), and both plots were generated with $\alpha_1=\alpha_2=n=1$, with $m_\phi = 4\,m_\psi$.}
\label{fig:n3}
\end{figure}

We have also considered a case in which both $m_\phi$ and $m_\psi$ are \emph{simultaneously} increasing with $\ell$, while their ratio is kept fixed. In Fig.~\ref{fig:n3}, we use $m_\phi = 4\,m_{\psi}$, and increase $m_\phi$, with $\ell=|m_\phi|+|m_\psi|$. Since both $m_\phi$ and $m_\psi$ are increasing, we expect the matched asymptotic expansion analysis to give a better approximation. We indeed see that this is the case: for $m_\psi=-1$ and $m_\phi=-4$, the matched asymptotic expansion result is barely discernible from the numerical data. Note that the colour coding in Fig.~\ref{fig:n3} is the same as in Fig.~\ref{fig:n1}.

\begin{figure}[h]
\centering
\includegraphics[width=0.9\linewidth]{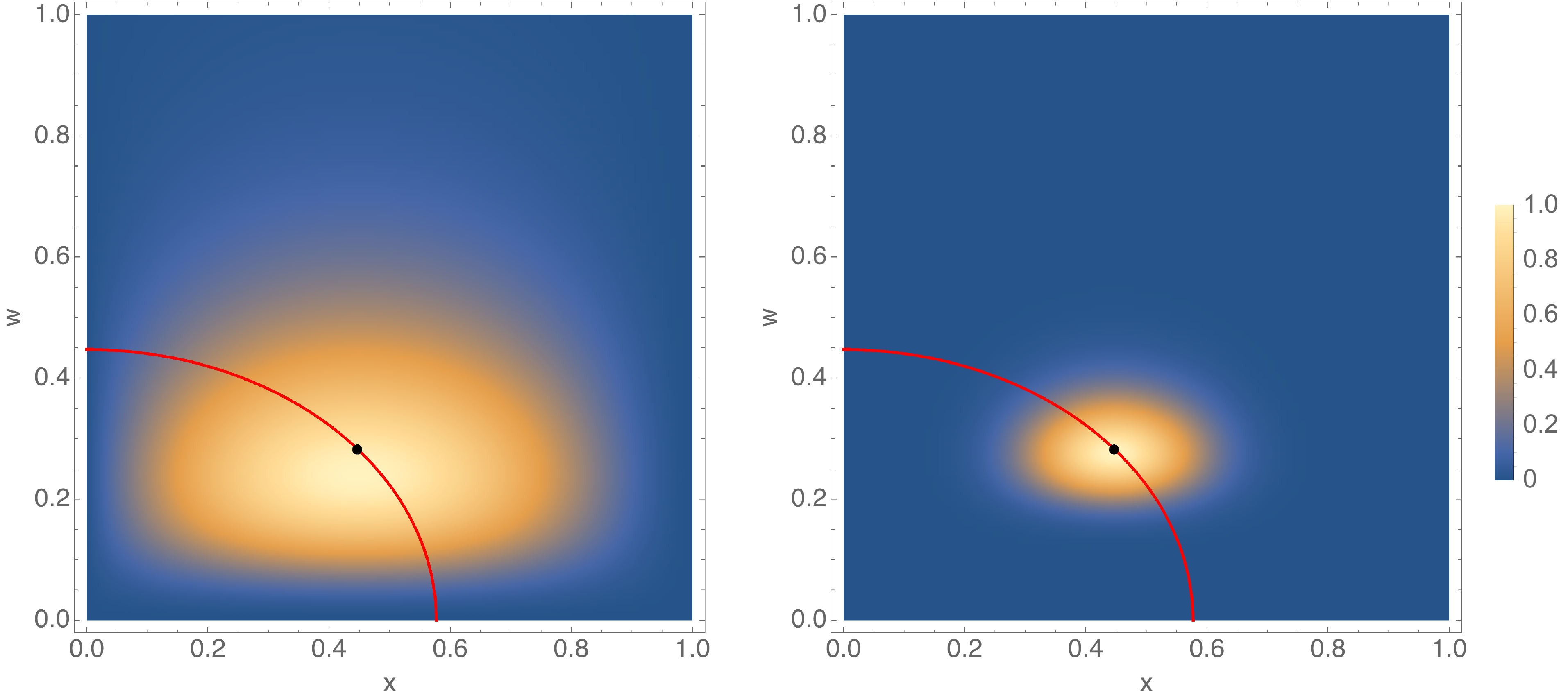}
\caption{Contour plot for $|\Phi|$ as a function of $w$ and $x$: on the \emph{left panel} we have $m_\phi=-4$ and on the \emph{right panel} we have $m_\phi=-32$. Both panels were generated with $\alpha_1=\alpha_2=n=1$ and $m_\phi = 4m_\psi$.}
\label{fig:n4}
\end{figure}

In Fig.~\ref{fig:n4}, we plot the normalised quasinormal mode as a function of $w$ and $x$, for the case $m_\phi = 4m_\psi$. As before, its peak is located exactly at the point predicted in section \ref{sec:geodesics}. Furthermore, the peak gets more and more sharp as we increase $\ell = |m_\phi|+|m_\psi|$.

\begin{figure}[h]
\centering
\includegraphics[width=0.4\linewidth]{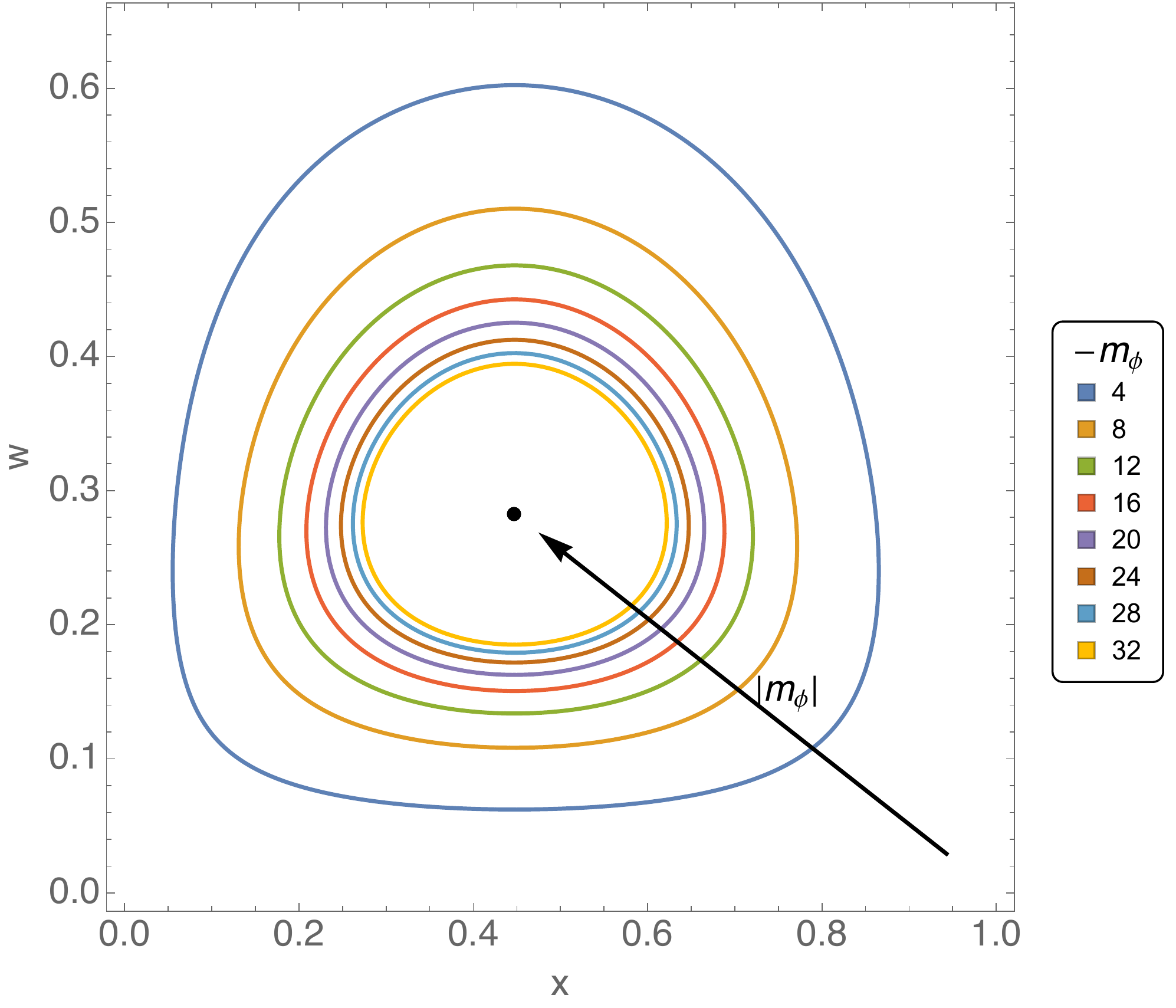}
\caption{Contour lines for $|\Phi|=1/4$ at fixed $u=t-r$, as a function of $w$ and $x$. All curves were generated with $\alpha_1=\alpha_2=n=1$ and $m_\phi=4\,m_\psi$.}
\label{fig:n5}
\end{figure}

One can go further, and determine the width of quasinormal mode around its maximum. It turns out to scale as $\sqrt{\ell}$, as expected from geometrical optics. This is best observed in Fig.~\ref{fig:n5}, where we plot the contour lines of $|\Phi|=1/5$, for several values of $m_{\phi}$. The arrow in the plot indicates the direction of increasing $(-m_\phi)$, and the point in the middle indicates the geometric optics prediction for the location of the maximum of $|\Phi|$.


\subsection{Lower bound on decay rate}

Proofs of nonlinear stability usually require first establishing {\it uniform} decay for linear perturbation. The first step is to establish decay of some {\it non-degenerate} energy. We consider some spacelike Cauchy surface $\Sigma_0$ and let $\Sigma_t$ denote the surface obtained by translation $\Sigma_0$ through parameter distance $t$ w.r.t. the Killing field $V$. A non-degenerate energy $E_1(t)$ is an integral over $\Sigma_t$ of some quantity quadratic in $\partial \Phi$, such that $E_1(t)$ is positive definite. Note that the {\it conserved} energy does not have this property because it degenerates on the evanescent ergosurface. 

Ideally one would like to establish a quantitative uniform energy decay result of the form 
\be
 E_1(t) \le g(t) E_1(0)
\ee
for some function $g(t)$, independent of $\Phi$, with $g(t) \rightarrow 0$ as $t \rightarrow \infty$. This is uniform because it applies to {\it any} perturbation $\Phi$ with $g$ independent of the perturbation. If $g(t)$ decays fast enough (e.g. $t^{-p}$ for large enough $p$) then one can hope to establish non-linear stability. However, when trapping is present, it is known that a decay result of this form does not exist \cite{Sbierski:2013mva}. Instead the best one can hope for is energy decay with "loss of a derivative", which means that one has
\be
\label{energydecay}
 E_1(t) \le g(t) E_2(0)
\ee
where $E_2(t)$ is a {\it second order} energy, i.e., a positive functional of $\partial \Phi$ {\it and} $\partial^2 \Phi$ defined as an integral over $\Sigma_t$. For example, the Schwarzschild solution, which exhibits unstable trapping at the photon sphere, admits a result of the above form with $g(t) \propto t^{-2}$ \cite{Dafermos:2008en}. 

Energy-decay results of the above form have also been obtained for spacetimes with stable trapping, but the function $g(t)$ decays very slowly. For AdS black holes \cite{Holzegel:2013kna}, and also for ultracompact neutron stars \cite{Keir:2014oka}, results of the form (\ref{energydecay}) have been proved with $g(t) =  (\log (2+t) )^{-2}$. Moreover, in both of these examples, this result is sharp in the sense that if one picks $g(t)$ decaying faster than this then one can construct solutions which violate (\ref{energydecay}). In both cases, one can also obtain pointwise decay results for the field $\Phi$.

We can now use our quasinormal modes to show that the decay is evern slower for the supersymmetric microstate geometries studied above. Quasinormal modes do not have finite energy when defined on a surface of constant $t$ in the coordinates of (\ref{eq:metric}). This is because such modes diverge at spatial infinity. However, it is well known that quasinormal modes are finite at future null infinity. Therefore we will pick our Cauchy surfaces $\Sigma_0$ to extend to future null infinity. 

Now consider a quasinormal mode with large $\ell$. Since $E_2$ is quadratic in second derivatives of $\Phi$, we expect $E_2(0) < C \ell^4$ for some $C>0$ independent of $\ell$. Hence if (\ref{energydecay}) holds we must have $E_1(t) < C \ell^4 g(t)$. On the LHS we have
\be
 E_1(t) \sim \ell^2 e^{2 \omega_I t} 
\ee
where $\omega = \omega_R + i \omega_I$. The factor of $\ell^2$ comes from the fact that $E_1$ is quadratic in first derivatives of $\Phi$. More precisely, we can find some constant $D>0$, independent of $\ell$, such that 
\be
 E_1(t) > D \ell^2 e^{2 \omega_I t} 
\ee 
hence if (\ref{energydecay}) holds then we must have
\be
 D e^{2 \omega_I t} < C \ell^2 g(t)
\ee
For example, consider $g(t) = (\log (2+t))^{-2}$ as for the examples discussed above. Set $t = e^{\alpha \ell}$ for some $\alpha>0$. Then we need (using our result for $\omega_I$) 
\be
 D \exp ( -2\beta  e^{-2\ell \log \ell} e^{\alpha \ell} ) \stackrel{<}{\sim} \frac{C}{\alpha^2}  
\ee
where $\beta>0$ is the coefficient in our large $\ell$ expression for $\omega_I$ derived above. Now taking the limit $\ell \rightarrow \infty$ gives $D \stackrel{<}{\sim} C/\alpha^2$, which we can violate by taking $\alpha$ large enough. This proves that a uniform decay result of the form (\ref{energydecay}) {\it cannot} exist with $g(t) = (\log (2+t))^{-2}$, so the decay in a supersymmetric microstate geometry is slower than for an AdS black hole or an ultracompact neutron star. 

An example of a function $g(t)$ for which our quasinormal modes are consistent with (\ref{energydecay}) is given by
\be
\label{gdecay}
 g(t) = \ell^{-2} \qquad {\rm where} \qquad  2 \ell \log \ell = \log (2+t) \qquad {\rm for} \qquad \ell \gg 1
\ee
Of course, we are not claiming that a result of the form (\ref{energydecay}) exists, merely that it is not ruled out by the behaviour of quasinormal modes. Such decay is much too slow to be of any use in establishing {\it nonlinear} stability.

The above analysis can be made rigorous by replacing quasinormal modes with {\it quasimodes}. These are {\it approximate} solutions of the wave equation which are compactly supported. In particular, they vanish in a neighbourhood at spatial infinity so one can work with a foliation of constant coordinate time $t$ in the coordinates of (\ref{eq:metric}) so the surfaces $\Sigma_t$ extend to spatial infinity. Using quasimodes one can prove the following \cite{joepaper}
\begin{theorem}
\label{theorem slow decay}
 Let $k_1$, $k_2 > 0$. Let $\ell$ satisfy the following equation:
\begin{equation}
\label{equation ell log ell}
 \ell \log \ell = \log(2 + t)
\end{equation}
Then there exists a universal positive constant $C_{k_1,k_2} > 0$ such that the following holds: for solutions $\Phi$ to the linear wave equation $\Box_g \Phi = 0$,
\begin{equation}
 \limsup_{t \rightarrow \infty} \, \sup_{\Phi \in H^{k_1 + k_2}(\Sigma_0)} \frac{||\Phi||^2_{H^{k_1}(\Sigma_t)}}{||\Phi||^2_{H^{k_1 + k_2}(\Sigma_0)}} \, \ell^{2k_2}  \geq C_{k_1,k_2}
\end{equation}
In particular, for any $k_1$, $k_2$ this gives sub-polynomial decay.
\end{theorem}
Here $||\Phi||^2_{H^{k_1}(\Sigma_t)}$ is the $k$th Sobolev norm associated to $\Sigma_t$, i.e., the norm involving an integral over $\Sigma_t$ of the sum of squares of the first $k$ derivatives of $\Phi$. Our heuristic argument above corresponds to the case $k_1 = k_2 = 1$ of this theorem. In general, the theorem allows for a loss of $k_2$ derivatives. 

\subsection*{Acknowledgments}

We have enjoyed numerous helpful discussions with Joe Keir, who will present a mathematically rigorous version of the slow decay result in Ref. \cite{joepaper}. We are also grateful to Mihalis Dafermos, Oscar Dias, Gary Horowitz, David Turton and especially Don Marolf for useful discussions. This work was supported by European Research Council grant ERC-2011-StG 279363-HiDGR.

\appendix

\section{Appendix: 2-charge microstate geometries}
\label{app:2charge}

\subsection{The metric}

We consider the 2-charge supersymmetric microstate geometries constructed in \cite{maldacena2000}. These are obtained by setting $Q_p=0$ in the solution described in section \ref{sec:3charge}. Ref. \cite{lunin2002} describes a whole family of such solutions, but we will only consider the maximally rotating solution with a circular profile. The metric for this 2-charge $D1-D5$ microstate geometry (in the form given in \cite{Lunin:2001}) is \begin{equation}\begin{split} ds^2=& -\frac{1}{h}(dt^2-dz^2)+hf\left(d\theta^2+\frac{dr^2}{r^2+a^2}\right)-\frac{2a \sqrt{\Qo\Qf}}{hf}\left(\cos^2\theta dz d\psi+\sin^2\theta dt d\phi\right) \\& + h\Big[\left(r^2+\frac{a^2\Qo\Qf\cos^2\theta}{h^2f^2}\right)\cos^2\theta d\psi^2+\left(r^2+a^2-\frac{a^2\Qo\Qf\sin^2\theta}{h^2f^2}\right)\sin^2\theta d\phi^2\Big] \label{eq:2metric} \end{split}\end{equation} where \begin{equation}
f=r^2+a^2\cos^2\theta,\;\; h=\Big[\left(1+\frac{\Qo}{f}\right)\left(1+\frac{\Qf}{f}\right)\Big]^{1/2}
\end{equation} and $a$ is defined in \eqref{eq:a}.

\subsection{Evanescent ergosurface}
As in the 3-charge microstate geometry, the globally null Killing vector field is \begin{equation}
V=\frac{\pd}{\pd t}+\frac{\pd}{\pd z}
\end{equation} and the evanescent ergosurface $\mathcal{S}_2$  defined by $V\cdot Z=0$ is at $f=0$; this is defined by $r=0$ and $\theta=\pi/2$.

In the 2-charge geometry the Kaluza-Klein circle pinches off smoothly at $f=0$ \cite{lunin2002}. The $\psi-$direction also shrinks to zero size at $f=0$ (in the same way as at the origin of polar coordinates) so that at constant $t$, $\mathcal{S}_2$ has topology $S^1$ where the coordinate around this circle is $\phi$. There are several differences between the evanescent ergosurface in the 2- and 3-charge geometries. First of all they have different dimensions: the 3-charge $\mathcal{S}$ is 5 dimensional whilst the 2-charge $\mathcal{S}_2$ is only 2 dimensional. In the 2-charge case the Killing vector field $T=\partial/\partial t$ is timelike everywhere except on $\mathcal{S}_2$ where it is null ($V$ is null everywhere and $Z= \partial/\partial z$ vanishes on $\mathcal{S}_2$) and so in this case there is no ergoregion, in contrast with the 3-charge case where $T$ is spacelike on $\mathcal{S}$. 

There are zero energy null geodesics with tangent vector $V$ which are stably trapped on $\mathcal{S}_2$ and thus stay at constant $r=0,\,\theta=\pi/2$. In the same way as for the 3-charge geometry this follows from the discussion in section \ref{sec:geodesics}.

\subsection{2-charge quasinormal modes}
The wave equation separates in the 2-charge microstate geometries in the coordinates of \eqref{eq:2metric} (see ref. \cite{Lunin:2001}) in the same way as for the 3-charge geometry but with $n=0$. In the wave equation we will therefore again use the ansatz \begin{equation}
\Phi(t,z,r,\theta,\phi,\psi)=e^{-i\omega t+i\lambda z+i\mpsi \psi+i\mphi \phi}\Phi_r(r)\Phi_{\theta}(\theta).
\end{equation}
However, if we are looking for modes that correspond, via the geometric optics approximation, to null geodesics with tangent vector $V$ that are stably trapped on $\mathcal{S}_2$ we must set $\mpsi=0$ because the corresponding geodesics are localized at $\theta=\pi/2$ so they have $\ppsi=0$.

Ref. \cite{Lunin:2001} discusses scattering solutions of the wave equation with low frequencies. Here we will find quasinormal modes with $|\mphi|\gg 1$. As for the 3-charge case, we look specifically for solutions where $\wb,\lb=O(1)\ll |\mphi|$, motivated by the geometric optics approximation since the geodesics with tangent $V$ on $\mathcal{S}_2$ have zero energy and Kaluza-Klein momentum.  

\subsubsection{2-charge matched asymptotic expansion}
After separating variables, the equation for $\Phi_{\theta}(\theta)$ is exactly the same as \eqref{eq:theq} with $\mpsi=0,\,n=0$ and $\eta=1$. Note that $\mpsi=0$ implies that $ j=0$ and that if we write \eqref{eq:theq} in the form of a Schr\"{o}dinger equation the potential is not strictly positive at $\theta=\pi/2$ on $\mathcal{S}_2$ so we have an 'allowed' region there. 

Exactly as in section \ref{sec:matching}, from eq. \eqref{eq:theq} the separation constant is $A=\ell(\ell+2)+O(1)$ where \begin{equation}
\ell \geq|\mphi|,\;\ell \in \mathbb{Z}.
\end{equation} We will construct quasinormal modes satisfying $\ell \gg 1$ and $|\mphi|=O(\ell)$. 

The differences to the calculation for the 3-charge case arise in the radial equation. We still have equation \eqref{eq:yeq} for $\Psi_r(y)$ but there are important differences in the coefficients $b$ and $c$: \begin{equation} \begin{split}
b_0=s^2(m^2-1&),\; b_1=-2s^2(1+m \wb) \\
c_0=0=c_1& \Rightarrow c=c_2=\alpha^4\lb^2.
\end{split}
\end{equation}

From the calculation for the 3-charge case we expect that we will have to set $m=-1$; in this case $b_0=0$ and $b=O(\ell)$. When we define each region we will allow either $b_0=0$ or $b_0\neq 0$ and use (assuming $\tilde{\kappa}=O(1)$): \begin{enumerate}[1)]
	\item $y\ll \ell^{\frac{1}{4}}$: $\tilde{\kappa}^2y^6\ll: V(y)\approx ay^4-by^2+c$
	\item $1\ll y \ll \ell$: $V(y)\approx a y^4$;
	\item $y\gg\sqrt{\ell}$: $\ell ^2 (y^2+C)\ll \tilde{\kappa}^2y^6$ and $V(y)\approx -\tilde{\kappa}^2y^6+a y^4$.
\end{enumerate}
Although the regions themselves are slightly different to those used in the 3-charge case, region 2 still overlaps both regions 1 and 3 and we approximate the equation in the same way as before in each region.

Therefore the analysis of section \ref{sec:3charge} follows through in exactly the same way as before; the fact that $c=O(1)$ doesn't change anything in the method or matching and we reach the same conditions as in the 3-charge case.

First of all, substituting $j=0$ and $n=0$ into equation \eqref{eq:lcancel}, the requirement that the frequencies do not scale with $\ell$, implies that \begin{equation}
m=-1
\end{equation} as we anticipated so that we do indeed have $b=O(\ell)$.

For the real part of the frequency we substitute $\eta=1$ into \eqref{eq:wr} (or substitute $c_2$ and the other necessary values into \eqref{eq:pole}) to find that at leading order \begin{equation}
\wb_R=2(N+1)+\lb. \label{eq:w2r}
\end{equation}

For the imaginary part of the frequency given in \eqref{eq:wi2} we set $\mu=0$ to find \begin{equation}
\wb_I=-D_2 s \tilde{\kappa}_0^2 e^{-2\ell\log\ell+\ell\left(2+2\log\frac{\tilde{\kappa}_0\sqrt{\alpha}}{2}\right)+(N-\frac{3}{2})\log \ell+O(1)} \label{eq:w2i}
\end{equation} for some positive constant $D_2$ and $\tilde{\kappa}_0=\sqrt{\wb_{R,0}^2-\lambda^2}$ where $\wb_{R,0}$ is the real part of $\wb$ calculated to leading order only in \eqref{eq:w2r}.

In both the 2- and 3-charge geometries the imaginary part of $\omega$ is negative and  $O(e^{-2\ell\log \ell})$ as $\ell\rightarrow\infty$ when $\tilde{\kappa}=O(1)$. Hence the dimension of the evanescent ergosurface does not seem to make a difference to the rate at which the modes decay. 

\subsubsection{2-charge quasinormal mode frequencies scaling with $\ell$}

The angular equation for the 2-charge case is exactly the same as in the 3-charge case, but we had to modify the calculation of section \ref{sec:matching} because some of the coefficients in the potential for the radial equation were zero at leading order. However, if we now assume that $|\wb-\lb|=O(1)$ but $\lb=O(\ell)$ so that the frequency scales with $\ell$, the coefficients in the potential are non-zero at leading order and the calculation for the quasinormal frequencies that scale with $\ell$ is exactly the same as in section \ref{sec:matchingell}.

To obtain the quasinormal modes for the 2-charge case from the 3-charge calculation we set $n=0$. Previously we also had to set $j=0$ because we were looking for quasinormal modes localised near null geodesics stably trapped on $\mathcal{S}_2$. Now we want to find solutions of the wave equation localised near null geodesics that are  stably trapped away from $\mathcal{S}_2$; these do not necessarily have $j=0$. However, in the calculation of section \ref{sec:matchingell} we assume that we still have $n-(n+1)j\geq 0$; for ease of calculation we will therefore still assume that $j=0$ here so we are looking for solutions localised near $\theta=\pi/2$ but not on $\mathcal{S}_2$. 

In this case we can simply substitute $n=0$ and $j=0$ into the results of section \ref{sec:matchingell}. We find the real and imaginary parts of the quasinormal frequencies from equations \eqref{eq:wr2} and \eqref{eq:wi2l} respectively: at leading order

\begin{equation}
\wb_R=\lb+\frac{2\eta}{P'}(N+1) +O(\ell^{-1}) \label{eq:wrl2charge}
\end{equation} where 
\begin{equation}
P'=1+\frac{\lb\alpha^2}{\ell}\left(1-\frac{\ell}{\lb+\ell}\right)+\frac{\lb}{\ell}\frac{\Qo+\Qf}{R_z^2}.
\end{equation}
If we define
\begin{equation*}
\mu''=\frac{\lb}{\ell}>0
\end{equation*}
we find that the imaginary part in the limit $\ell\rightarrow\infty$ is 
\begin{equation}
\wb_I=-D_2'e^{-\ell\log\ell+\ell\left[2-\mu''\log\mu''+(1+\mu'')\log(\mu''+1)+2\log\frac{\tilde{\kappa}_0\sqrt{\alpha}}{2\sqrt{\ell}}\right]+(N+\frac{1}{2}-\nu_1)\log \ell} +O(l^{-1}) \label{eq:wil2charge}
\end{equation} where $\nu_1$ is independent of $\ell$ and defined in \eqref{eq:bcn} with $n=0$ and $j=0$, $D_2'$ is a constant proportional to $s^{\nu_1}$ that vanishes in the decoupling limit and $\tilde{\kappa}_0=\sqrt{\wb^2_R-\lb^2}$ with $\wb_R$ defined in \eqref{eq:wrl2charge}.

\section{Appendix: Quasimode construction}
\label{app:quasi}

Quasimodes are {\it approximate} solutions of the wave equation, with exponentially small error \cite{Holzegel:2013kna,Keir:2014oka}. Quasimodes can be used to study local features of potentials, and establish rigorous lower bounds on the uniform decay of fields. Even though one can envisage such a construction for generic backgrounds, it has only been firmly established for backgrounds that admit separable solutions \cite{Holzegel:2013kna,Keir:2014oka}. In such cases, the equations of motion governing how certain perturbations propagate on such backgrounds, become a set of coupled ordinary differential equations, for which potentials can be defined. Our geometries fall into that class.

Quasimodes are constructed as follows. One first restricts to a {\it finite} domain and impose boundary conditions at the edges of this domain. We choose our boundary conditions to be such that at the centre, $w=0$, the quasimode is regular, and at a given radius, say $w=w_c$ we impose a Dirichlet boundary condition $\Phi=0$.  The choice of $w_c$ is largely irrelevant, except we want to make it sufficiently large that any interesting feature in our potential lies in the interval $w\in(0,w_c)$. We solve this Dirichlet problem for $w < w_c$, which gives a set of normal mode frequencies, and then set $\Phi=0$ for $w>w_c$. The resulting solutions are not smooth at $w=w_c$; one defines quasimodes by applying a smoothing procedure near $w=w_c$, which means that one no longer has an exact solution to the wave equation: there is a small error near $w_c$. 

We will perform the first part of this construction, i.e., solution of the Dirichlet problem. It turns out that the associated normal mode frequencies give an excellent fit to the real part of the frequences of our quasinormal modes. For the sake of presentation, we will only describe below the case in which we kept $m_\psi$ fixed, but allow $m_\phi$ to become arbitrarily large. In addition, we will set $\ell = |m_\phi|+|m_\psi|$. 

The idea is simple, we start with a consistent ansatz for the angular and radial eigenfunctions and eigenvalues. These take the following form:
\begin{align}
X(x)=x^{|m_{\psi}|}(1-x^2)^{\frac{|m_\phi|}{2}}\,\sum_{k=0}^{+\infty}\frac{\widetilde{X}_k(x)}{|m_\phi|^k}\,,\qquad W(w)=e^{-|m_\phi|\widetilde{\phi}(w)}W_0(w)\left[1+\sum_{k=1}^{+\infty}\frac{\widetilde{W}_k(w)}{|m_\phi|^k}\right]\,,\nonumber
\\
A=(|m_\phi|+|m_\psi|)(|m_\phi|+|m_\psi|+2)+\sum_{k=0}^{+\infty}\frac{\widetilde{A}_k}{|m_\phi|^k}\,,\qquad\text{and}\qquad \tilde{\omega}=\sum_{k=0}^{+\infty}\frac{\varpi_k}{|m_\phi|^k}\,.\nonumber
\end{align}
Inputting these into the equations of motion, allows us to determine the coefficients
$$
\{\widetilde{X}_k(x),\widetilde{W}_k(w),\widetilde{A}_k,\varpi_k\}
$$
to any order in the expansion. For instance, keeping all parameters in the 3-charge microstate geometries gives
\begin{align}
\widetilde{A}_0=\alpha _1\,\alpha _2\,\eta \,n\,\varpi _0^2\,,\qquad \widetilde{A}_1 = -\alpha _1\,\alpha _2\,\eta\,\varpi _0 \left[(2 n+1) \varpi _0 \left(|m_{\psi }|+1\right)-2 n \varpi _1\right]\,,\nonumber
\\
\varpi_0 = 2\,\eta\,,\qquad \varpi_1 = -\frac{2 \left(\alpha _1+\alpha _2+\alpha _1 \alpha _2 n^2+\alpha _1 \alpha _2 n\right)}{\left[\left(\alpha _1+\alpha _2\right) n^2+\left(\alpha _1+\alpha _2\right)n+1\right]^3}\nonumber\,.
\end{align}

It is possible to go to higher orders in $k$, but the expressions become increasingly complicated. Progress can be made by choosing specific values for $\alpha_1$, $\alpha_2$, $n$ and $m_\psi$. For instance, for $\alpha_1=\alpha_2=n=-m_\psi=1$ (the parameters of Fig.~(\ref{fig:n1})), we find
$$
\tilde{\omega} = \frac{2}{5}-\frac{8}{125 \left| m_{\phi }\right|}+\frac{424}{3125 \left| m_{\phi }\right|^2}-\frac{21284}{78125 \left| m_{\phi}\right|^3}+\frac{968684}{1953125 \left| m_{\phi }\right|^4}-\frac{34114268}{48828125 \left| m_{\phi }\right|^5}+O\left(\left| m_{\phi }\right|^{-6}\right)\,.
$$


\begin{thebibliography}{99}

\bibitem{Lunin:2001jy} 
  O.~Lunin and S.~D.~Mathur,
  Nucl.\ Phys.\ B {\bf 623}, 342 (2002)
  doi:10.1016/S0550-3213(01)00620-4
  [hep-th/0109154].

\bibitem{maldacena2000}
J.~M.~Maldacena and L.~Maoz,
  JHEP {\bf 0212} (2002) 055
  [hep-th/0012025].

\bibitem{balasubramanian2000}
  V.~Balasubramanian, J.~de Boer, E.~Keski-Vakkuri and S.~F.~Ross,
  Phys.\ Rev.\ D {\bf 64} (2001) 064011
  [hep-th/0011217].


\bibitem{lunin2002}
O.~Lunin, J.~M.~Maldacena and L.~Maoz,
  hep-th/0212210.

\bibitem{Lunin:2004uu} 
  O.~Lunin,
  JHEP {\bf 0404}, 054 (2004)
  doi:10.1088/1126-6708/2004/04/054
  [hep-th/0404006].

\bibitem{Giusto:2004id} 
  S.~Giusto, S.~D.~Mathur and A.~Saxena,
  Nucl.\ Phys.\ B {\bf 701}, 357 (2004)
  [hep-th/0405017].


\bibitem{Giusto:2004ip} 
  S.~Giusto, S.~D.~Mathur and A.~Saxena,
  Nucl.\ Phys.\ B {\bf 710}, 425 (2005)
  [hep-th/0406103].

\bibitem{Giusto:2004kj} 
  S.~Giusto and S.~D.~Mathur,
  Nucl.\ Phys.\ B {\bf 729}, 203 (2005)
  doi:10.1016/j.nuclphysb.2005.09.037
  [hep-th/0409067].

\bibitem{Bena:2005va} 
  I.~Bena and N.~P.~Warner,
  Phys.\ Rev.\ D {\bf 74}, 066001 (2006)
  doi:10.1103/PhysRevD.74.066001
  [hep-th/0505166].

\bibitem{Berglund:2005vb} 
  P.~Berglund, E.~G.~Gimon and T.~S.~Levi,
  JHEP {\bf 0606}, 007 (2006)
  doi:10.1088/1126-6708/2006/06/007
  [hep-th/0505167].

\bibitem{Gibbons:2013tqa} 
  G.~W.~Gibbons and N.~P.~Warner,
  Class.\ Quant.\ Grav.\  {\bf 31}, 025016 (2014)
  doi:10.1088/0264-9381/31/2/025016
  [arXiv:1305.0957 [hep-th]].

\bibitem{Jejjala:2005yu} 
  V.~Jejjala, O.~Madden, S.~F.~Ross and G.~Titchener,
  Phys.\ Rev.\ D {\bf 71}, 124030 (2005)
  [hep-th/0504181].

\bibitem{Cardoso:2005gj} 
  V.~Cardoso, O.~J.~C.~Dias, J.~L.~Hovdebo and R.~C.~Myers,
  Phys.\ Rev.\ D {\bf 73}, 064031 (2006)
  [hep-th/0512277].

\bibitem{Breckenridge:1996is} 
  J.~C.~Breckenridge, R.~C.~Myers, A.~W.~Peet and C.~Vafa,
  Phys.\ Lett.\ B {\bf 391}, 93 (1997)
  doi:10.1016/S0370-2693(96)01460-8
  [hep-th/9602065].

\bibitem{Elvang:2004rt} 
  H.~Elvang, R.~Emparan, D.~Mateos and H.~S.~Reall,
  Phys.\ Rev.\ Lett.\  {\bf 93}, 211302 (2004)
  doi:10.1103/PhysRevLett.93.211302
  [hep-th/0407065].

\bibitem{Christodoulou:1993uv} 
  D.~Christodoulou and S.~Klainerman,
  Princeton University Press, Princeton, 1993


\bibitem{dafermos}
M. Dafermos and G. Holzegel, ``Dynamic instability of solitons in 4+1 dimensional gravity with negative cosmological constant," unpublished (2006). Available at: www.dpmms.cam.ac.uk/$\sim$md384/ADSinstability.pdf

\bibitem{Bizon:2011gg} 
  P.~Bizon and A.~Rostworowski,
  Phys.\ Rev.\ Lett.\  {\bf 107}, 031102 (2011)
  doi:10.1103/PhysRevLett.107.031102
  [arXiv:1104.3702 [gr-qc]].


\bibitem{Cardoso:2007ws} 
  V.~Cardoso, O.~J.~C.~Dias and R.~C.~Myers,
  Phys.\ Rev.\ D {\bf 76}, 105015 (2007)
  [arXiv:0707.3406 [hep-th]].

\bibitem{Sbierski:2013mva} 
  J.~Sbierski,
  arXiv:1311.2477 [math.AP].


\bibitem{Holzegel:2013kna} 
  G.~Holzegel and J.~Smulevici,
  arXiv:1303.5944 [gr-qc].

\bibitem{Keir:2014oka} 
  J.~Keir,
  Class.\ Quant.\ Grav.\  {\bf 33}, no. 13, 135009 (2016)
  doi:10.1088/0264-9381/33/13/135009
  [arXiv:1404.7036 [gr-qc]].

\bibitem{John:1981}
  F.~John,
  Commun.\ Pure Appl.\ Math.\ {\bf 34} 29 (1981),
  doi:10.1002/cpa.3160340103.


\bibitem{Klainerman:1984}
  S.~Klainerman,
  In ``Nonlinear systems of partial differential equations in applied mathematics," Part 1 (Santa Fe, N.M., 1984), volume 23
of Lectures in Appl. Math., pages 293–326. Amer.\ Math.\ Soc.\ , Providence, RI, 1986.

\bibitem{Lindblad:2004ue}
  H.~Lindblad and I.~Rodnianski,
  math/0411109 [math-ap].

\bibitem{Lindblad:2008}
  H.~Lindblad,
  Amer.\ J.\ Math.\ {\bf 130} (2008), no. 1, 115-157
  math/0511461 [math-ap].

\bibitem{Dias:2012tq} 
  O.~J.~C.~Dias, G.~T.~Horowitz, D.~Marolf and J.~E.~Santos,
  Class.\ Quant.\ Grav.\  {\bf 29}, 235019 (2012)
  doi:10.1088/0264-9381/29/23/235019
  [arXiv:1208.5772 [gr-qc]].

\bibitem{Festuccia:2008zx} 
  G.~Festuccia and H.~Liu,
  Adv.\ Sci.\ Lett.\  {\bf 2}, 221 (2009)
  doi:10.1166/asl.2009.1029
  [arXiv:0811.1033 [gr-qc]].



\bibitem{Gannot:2012pb} 
  O.~Gannot,
  Commun.\ Math.\ Phys.\  {\bf 330}, 771 (2014)
  doi:10.1007/s00220-014-2002-4
  [arXiv:1212.1907 [math.SP]].

\bibitem{Cardoso:2014sna} 
  V.~Cardoso, L.~C.~B.~Crispino, C.~F.~B.~Macedo, H.~Okawa and P.~Pani,
  Phys.\ Rev.\ D {\bf 90}, no. 4, 044069 (2014)
  doi:10.1103/PhysRevD.90.044069
  [arXiv:1406.5510 [gr-qc]].

\bibitem{joepaper}
J. Keir, arXiv:1609.01733 [gr-qc].

\bibitem{Mathur:2005zp} 
  S.~D.~Mathur,
  Fortsch.\ Phys.\  {\bf 53}, 793 (2005)
  doi:10.1002/prop.200410203
  [hep-th/0502050].

\bibitem{Aretakis:2011ha} 
  S.~Aretakis,
  Commun.\ Math.\ Phys.\  {\bf 307}, 17 (2011)
  doi:10.1007/s00220-011-1254-5
  [arXiv:1110.2007 [gr-qc]].

\bibitem{Aretakis:2011hc} 
  S.~Aretakis,
  Annales Henri Poincare {\bf 12}, 1491 (2011)
  doi:10.1007/s00023-011-0110-7
  [arXiv:1110.2009 [gr-qc]].





\bibitem{Kunduri:2014kja} 
  H.~K.~Kunduri and J.~Lucietti,
  Phys.\ Rev.\ Lett.\  {\bf 113}, no. 21, 211101 (2014)
  doi:10.1103/PhysRevLett.113.211101
  [arXiv:1408.6083 [hep-th]].

\bibitem{Tomizawa:2016kjh} 
  S.~Tomizawa and M.~Nozawa,
  arXiv:1606.06643 [hep-th].

\bibitem{Kunduri:2014iga} 
  H.~K.~Kunduri and J.~Lucietti,
  JHEP {\bf 1410}, 082 (2014)
  doi:10.1007/JHEP10(2014)082
  [arXiv:1407.8002 [hep-th]].

\bibitem{Giusto:2013rxa} 
  S.~Giusto, L.~Martucci, M.~Petrini and R.~Russo,
  Nucl.\ Phys.\ B {\bf 876}, 509 (2013)
  doi:10.1016/j.nuclphysb.2013.08.018
  [arXiv:1306.1745 [hep-th]].

\bibitem{Bena:2016ypk} 
  I.~Bena, S.~Giusto, E.~J.~Martinec, R.~Russo, M.~Shigemori, D.~Turton and N.~P.~Warner,
  arXiv:1607.03908 [hep-th].

\bibitem{Gibbons:1993xt} 
  G.~W.~Gibbons, D.~Kastor, L.~A.~J.~London, P.~K.~Townsend and J.~H.~Traschen,
  Nucl.\ Phys.\ B {\bf 416}, 850 (1994)
  doi:10.1016/0550-3213(94)90558-4
  [hep-th/9310118].


\bibitem{Niehoff:2016gbi} 
  B.~E.~Niehoff and H.~S.~Reall,
  JHEP {\bf 1604}, 130 (2016)
  doi:10.1007/JHEP04(2016)130
  [arXiv:1601.01898 [hep-th]].

\bibitem{Gutowski:2003rg} 
  J.~B.~Gutowski, D.~Martelli and H.~S.~Reall,
  Class.\ Quant.\ Grav.\  {\bf 20}, 5049 (2003)
  doi:10.1088/0264-9381/20/23/008
  [hep-th/0306235].

\bibitem{wald}
R.M. Wald {\it General Relativity}, University of Chicago Press (1984). 



\bibitem{Dafermos:2009uq}
  M.~Dafermos and I.~Rodnianski,
  XVIth International Congress on Mathematical Physics, P. Exner   (ed.), World Scientific, London, 2009, pp. 421-433
  [arXiv:0910.4957 [math.AP]].


\bibitem{Dafermos:2013bua}
  M.~Dafermos, G.~Holzegel and I.~Rodnianski,
  arXiv:1306.5364 [gr-qc].

\bibitem{Dafermos:2005eh}
M.~Dafermos and I.~Rodnianski,
Commun.\ Pure Appl.\ Math.\ {\bf 62} (2009) 859
[gr-qc/0512119].


\bibitem{Dafermos:2008en} 
  M.~Dafermos and I.~Rodnianski,
  Clay Math.\ Proc.\  {\bf 17}, 97 (2013)
  [arXiv:0811.0354 [gr-qc]].


  \bibitem{Aretakis:2012ei} 
  S.~Aretakis,
  Adv.\ Theor.\ Math.\ Phys.\  {\bf 19}, 507 (2015)
  doi:10.4310/ATMP.2015.v19.n3.a1
  [arXiv:1206.6598 [gr-qc]].

\bibitem{Lucietti:2012sf} 
  J.~Lucietti and H.~S.~Reall,
  Phys.\ Rev.\ D {\bf 86}, 104030 (2012)
  doi:10.1103/PhysRevD.86.104030
  [arXiv:1208.1437 [gr-qc]].


\bibitem{Dafermos:2010hb}
M.~Dafermos and I.~Rodnianski,
arXiv:1010.5132 [gr-qc].



\bibitem{Chakrabarty:2015foa} 
  B.~Chakrabarty, D.~Turton and A.~Virmani,
  JHEP {\bf 1511}, 063 (2015)
  doi:10.1007/JHEP11(2015)063
  [arXiv:1508.01231 [hep-th]].

\bibitem{Ferrari:1984}
	V.~Ferrari and B.~Mashhoon,
	Phys.\ Rev.\ D {\bf 30}, 295 (1984).

\bibitem{Yang:2012he} 
  H.~Yang, D.~A.~Nichols, F.~Zhang, A.~Zimmerman, Z.~Zhang and Y.~Chen,
  Phys.\ Rev.\ D {\bf 86}, 104006 (2012)
  [arXiv:1207.4253 [gr-qc]].

\bibitem{AS}{M. Abramowitz and I.A. Stegun, \textit{Handbook of Mathematical Functions with Formulas, Graphs and Mathematical Tables}, United States Department of Commerce, National Bureau of Standards, 1964.}

\bibitem{Giusto:2012yz} 
  S.~Giusto, O.~Lunin, S.~D.~Mathur and D.~Turton,
  JHEP {\bf 1302}, 050 (2013)
  doi:10.1007/JHEP02(2013)050
  [arXiv:1211.0306 [hep-th]].

\bibitem{Cardoso:2013pza} 
  V.~Cardoso, O.~J.~C.~Dias, G.~S.~Hartnett, L.~Lehner and J.~E.~Santos,
  JHEP {\bf 1404}, 183 (2014)
  doi:10.1007/JHEP04(2014)183
  [arXiv:1312.5323 [hep-th]].

\bibitem{Lunin:2001}
O.~Lunin and S.~D.~Mathur
Nucl.\ Phys.\ {\bf B615}, 285 (2001)
[hep-th/0107113]


\end{thebibliography}
\end{document}